\documentclass{aa}  

\usepackage[varg]{txfonts}
\usepackage{amssymb}
\usepackage{natbib}
\bibpunct{(}{)}{;}{a}{}{,}
\usepackage{amsmath} 
   
\newcommand{\amm}{NH$_3$}

\newcommand{\met}{CH$_3$OH}

\newcommand{\kms}{km~s$^{-1}$}
\newcommand{\cmc}{cm$^{-3}$}
\newcommand{\cmq}{cm$^{-2}$}
\newcommand{\ms}{$M_{\odot}$}
\newcommand{\ls}{$L_{\odot}$}

\newcommand{\msyr}{$M_{\odot}$~yr$^{-1}$}

\newcommand{\pas}{$\rlap{.}^{\prime\prime}$}
\newcommand{\degree}{$^{\circ}$}

\newcommand{\tkin}{$T_{kin}$}
\newcommand{\trot}{$T_{rot}$}
\newcommand{\tex}{$T_{ex}$}

\begin{document}


\title{Hot Ammonia around O-type Young Stars}

 \subtitle{I. JVLA imaging of \amm\,(6,6) to (14,14) in NGC7538~IRS1}

 \titlerunning{Hot NH$_3$ around O-type Young Stars: I. NGC7538~IRS1}

  \author{ C. Goddi \inst{1}
\and Q. Zhang \inst{2}
\and L. Moscadelli  \inst{3}
}

\offprints{C. Goddi,\\\email{goddi@jive.nl}}

\institute{Joint Institute for VLBI in Europe, Postbox 2, 7990 AA Dwingeloo, The Netherlands  
\and Harvard-Smithsonian Center for Astrophysics, 60 Garden Street, Cambridge, MA 02138
\and INAF, Osservatorio Astrofisico di Arcetri, Largo E. Fermi 5, 50125 Firenze, Italy
}



\abstract
{The formation of massive (O-type) stars through the same accretion processes as low-mass stars appears problematic, 
mainly because of the feedback massive stars provide to the environment  which halts the accretion. 
In order to constrain theoretical models of high-mass star formation, 
observational signatures of mass accretion in O-type forming stars are desirable.  
 The high-mass star forming region NGC7538~IRS1   (distance=2.7~kpc) is an ideal target, 
because VLBI measurements of \met\,masers recently identified  a triple system of high-mass young stellar object (YSOs)   in the region:  IRS1a, IRS1b, and IRS1c. 
The first two YSOs appears to be surrounded by rotating disks. 
}
{We want to characterize physical conditions and kinematics of circumstellar molecular gas around O-type young stars. 
Sub-arcsecond resolution observations of highly-excited lines from high-density tracers are useful,  since these probe  the hottest and densest gas, which presumably is close to O-type forming stars, i.e. in disks and innermost portions of envelopes. }
{Using the Karl Jansky Very Large Array (JVLA), we have mapped the hot and dense molecular gas in the hot core 
associated with NGC7538~IRS1, with $\sim$0\pas2 angular resolution, 
in seven metastable (J=K) inversion transitions of ammonia (NH$_3$): (J,K)=(6,6), (7,7), (9,9), (10,10), (12,12), (13,13), and (14,14). 
 These lines arise from energy levels between $\sim$400~K and $\sim$1950~K above the ground state, 
and are observed in absorption against the HC-HII region associated with  NGC7538~IRS1. 
The CH$_3$OH $J_K$= 13$_2$-13$_1$  and CH$_3$CN (2-1) lines were also included in our spectral setup, but only the former was detected. 
We also obtained sensitive continuum maps at frequencies between 25 and 35 GHz. 
}
{For each transition, we produced resolved images  of total intensity and velocity field, as well as position-velocity diagrams. 
The intensity maps show that the \amm\,absorption follows closely the continuum emission. 
With a 500 AU linear resolution, we resolve the elongated North-South \amm\,structure into two compact components:  the main core and a southernmost component. 
 Previous observations of the radio continuum with a 0\pas08 (or 200 AU) resolution, resolved the compact core in two (northern and southern) components.  
The velocity maps of the compact core show  a clear velocity gradient in all lines, which is indicative of rotation.  
 It is possible that the rotation is not in an accretion disk but in a (circumbinary) envelope, containing $\sim$40~\ms\,(dynamical mass). 
The core in fact hosts a binary system of massive YSOs,   associated with the two (northern and southern) components of the radio continuum, 
which have a separation of about 500 AU and velocities around --59~\kms\, and --56.4~\kms, respectively.
The  southernmost component, separated by 1000 AU and resolved in our \amm\,maps (0\pas2 beamsize) from the core, 
is associated with a third massive YSO, with a  velocity around --60~\kms. 
These features correspond to the triple system of high-mass YSOs IRS1a, IRS1b, and IRS1c.
 In addition, we derive rotational temperatures, \amm\,column densities, H$_2$ gas densities, and  gas masses from the \amm\,data. 
Surprisingly, measurements of the hyperfine structure show total optical depths of 10-26 even for these highly-excited lines, 
among the largest found so far in hot \amm\,in high-mass star forming regions.  
From ratios of optical depths as well as rotational temperature diagrams of the observed ortho and para transitions, 
 we derive a rotational temperature $\sim 280$~K,  \amm\,column densities $\sim 1.4-2.5 \times 10^{19}$~cm$^{-2}$, 
 total H$_2$ volume densities $\sim 3.5-6.2 \times 10^{10}$~\cmc\,and a total gas mass in the range of 19--34 M$_\odot$ (the latter two assume [\amm]/[H$_2$]=10$^{-7}$), for the main core.
For the southern component, we derive a temperature of 110 K,  a molecular density of $\sim 0.7-2 \times 10^{10}$~\cmc\,
and a gas mass in the range of 4--12 M$_\odot$. 
}
{We conclude that NGC7538~IRS1 is the densest hot molecular core known to date, 
containing a rotating envelope which hosts a multiple system of high-mass YSOs, possibly surrounded by accretion disks. 
 Future JVLA observations in the A-configuration are needed to resolve the binary system in the core and may allow to study the gas kinematics in the accretion disks  associated with individual binary members. 
}

 
\maketitle

\section{Introduction}
\label{intro}

The standard theory of low-mass star formation predicts that protostars form from the gravitational collapse of molecular cores  and mass accretes onto the protostar via an accretion  disk  \citep[e.g.,][]{Shu87}. 
However, whether the formation of high-mass (O-B type) stars follows a similar process is still a matter of debate. 
The main difficulty arises because the intense radiation pressure from the stellar luminosity and the thermal pressure from the HII region 
around the massive young stellar objects (YSOs) may be sufficient to reverse the accretion flow and prevent matter from reaching the star. 
 Recent theoretical studies  have however demonstrated that the radiation pressure problem can be solved 
 if accretion occurs through a circumstellar disk continuously fed from an infalling envelope \citep[e.g.,][]{Kuiper10,Kuiper11}, 
 thus explaining the formation of stars up to 140 $M_\odot$.
Despite the theoretical evidence, while a handful of disk candidates in B-type protostars ($M<20$ \ms) has been reported in the literature 
in recent years \citep[][and references therein]{Mosca13,Alvaro13,Cesa07,Zhang05,Beuther05},
there have been no clear evidence for accretion disks around more massive O-type stars so far \citep{Qiu12}.  
Likewise, evidence for bulk infall motions has been presented only towards a few luminous star forming regions, 
mostly based on observations of inverse P-Cygni profiles of molecular lines 
along the line-of-sight to  a bright background (dust or HII emission), 
with redshifted absorption and blueshifted emission 
\citep{ZhangHo97, SollinsHo05, Beltran06, Beltran11}. 
The problem is that O-type stars are mostly found at large distances ($>$1 kpc) and deeply embedded inside dense massive cores (likely containing protoclusters), 
therefore confusion generally precludes establishing whether the globally infalling (and rotating) material in the clump,  
on scales of tens of thousands of AU,   
would feed a central  cluster of lower-mass protostars or accrete onto  a single O-type star \citep[e.g.,][]{SollinsHo05, Beltran11}. 
In order to resolve and "decouple" individual rotating  disks from the infalling envelopes, 
 infall and rotation should  be measured in the proximity ($<$1000 AU) to  individual protostars. 
 This has been done so far only in a few YSOs with intermediate-mass ($<$20~\ms) using molecular masers observed with Very Long Baseline Interferometry (VLBI),  
 which provide the gas 3D velocity field on AU scales \citep[e.g.,][]{Matthews10,Goddi11b}.

Therefore, observational signatures of mass accretion in O-type forming stars remain one of the main issues in observational high-mass star formation (HMSF).
From an observational point of view, two elements are key: the high angular resolution and the choice of the molecular tracer. 
In this context, sub-arcsecond resolution observations of highly-excited lines from high-density tracers are useful, as they enable to probe  the hottest and densest gas, which presumably lies at small radii  from O-type forming stars, i.e. around centrifugally-supported disks and/or infalling envelopes. 
Traditionally, accretion signatures have been searched using molecular transitions in the (sub)mm regime \citep[e.g.,][]{Beltran05}. 
These transitions however can only probe outer parts of disks/envelopes where dust optical depth is low, but cannot penetrate the innermost regions, where dust emission is optically thick and, hence, dominates over molecular emission at a given frequency. 
We argue instead that observations of optically thin, highly-excited molecular~lines at $\lambda\sim$1 cm  could be more effective than (sub)mm interferometry in probing the innermost regions of disks/envelopes around OB-type young stars. 

\subsection{JVLA Imaging of Highly Excited Inversion Transitions of \amm in High-mass Star Forming Regions}

\amm\, has a symmetric top structure which produces inversion doublets as the nitrogen nucleus oscillates through the plane of the hydrogen nuclei.
The frequencies of the transitions between two states of an
inversion-doublet depend on the total angular momentum $J$ and its 
projection on the molecular axis, $K$, and are therefore described by the pair of quantum numbers ($J$,$K$):  
$K$ = 0, 3, 6, 9, 12... defines {\it ortho}-NH$_3$ and $K$ = 1, 2, 4, 5, 7... defines {\it para}-NH$_3$. 
The non-metastable ($J \ne K$) levels decay downward quickly (A $\sim10^{-2}$ sec$^{-1}$ ), populating the metastable ($J=K$) levels. 
The metastable levels cannot radiatively decay and are collisionally excited, 
thus multiple transitions can be used to calculate the kinetic temperature (\tkin) 
using a rotational diagram analysis \citep{HoTownes83}. 
Therefore, NH$_3$ is an excellent "thermometer" of dense molecular gas. 
From a chemical viewpoint,  two basic chemical properties make \amm\,unique. 
Like other nitrogen bearing species (e.g. N$_2$H$^+$), it is not easily depleted and is therefore prominent in quiescent cloud cores. 
Moreover, with a dust evaporation temperature of about 100 K, 
hydrogenated nitrogen-bearing molecules like \amm, copiously synthesized in dust grain mantles, 
can be extremely abundant in the gas phase in the neighborhood of recently formed O-type stars.  
But perhaps the most appealing aspect of \amm\,among all is that it can trace excitation up to temperatures of $\sim $2000\,K 
by observing its inversion transitions within a relatively narrow frequency 
range, 20 -- 40\,GHz \citep[e.g.,][]{HoTownes83}.

  Observations of highly excited \amm\,inversion lines present multiple advantages. 
i) The high excitation temperatures  imply that these transitions trace the hottest molecular gas in the vicinity of the central YSO(s) rather than the cold gas in the circumstellar envelopes. 
ii) Molecular transitions in the cm-regime tend to be optically thin and can more easily probe the innermost   disk/envelope regions than the (sub)mm lines \citep[e.g.,][]{Krumholz07}. 
iii) Measuring infall is easier in cm lines than in the mm regime, because the molecular gas can be seen in absorption against the bright continuum background of H II regions, 
and the absorption can be sufficiently strong even when probed with high angular resolution 
(the same lines in emission would be submerged by noise),
 thus enabling  studies of gas kinematics at very small scales.  
iv)  The detection of faint lines of high-density tracers in the (sub)mm spectrum can be severely affected by blending with stronger transitions from  other species (i.e. line forest); this is not a problem at radio wavelengths. 
 
 A first attempt in this direction was a study conducted in the southern hemisphere with the Australia Telescope Compact Array (ATCA) in a small sample of HMSF regions \citep{BeutherWalsh07,BeutherWalsh08,BeutherWalsh09}. 
 This study revealed clear signatures of rotation and/or infall motions in about half of the observed sample,  
but did not identify Keplerian signatures or even flattened structures resembling accretion disks in any of the sources, at least on the linear scales probed by ATCA ($10^3-10^4$ AU). 
 Besides the limited angular resolution (typically 0\pas5-2\arcsec), the ATCA study was
  limited to just two lines, the (4,4) and (5,5), of modest excitation ($E_u<$400 K),  
which turned out to have high optical depths, preventing reasonable rotational temperature estimates. 
A step forward was taken by \citet{Goddi11a},  who used the JVLA in C-configuration (0\pas7 beamsize, 300 AU) to image for the first time the \amm~lines (6,6) to (12,12) (with $E_u=$400-1500 K) in the Orion Hot Core. 
The multi-transition data set enabled to produce resolved images of the integrated intensity, velocity field, rotational temperature and column density of \amm. 
This in turn enabled  small-scale investigation of kinematics  and physical conditions of hot gas, 
and eventually identification of sources responsible for heating the famous Orion Hot Core \citep{Goddi11a}.  
 
We have started a program with the  Karl Jansky Very Large Array (JVLA) at $\lambda\sim$1 cm to observe  highly-excited inversion lines of \amm~toward Galactic hot molecular cores.
The sample has been selected from a survey conducted with the Effelsberg 100-m telescope \citep{Mauersberger86}. 
All the selected targets are among the most luminous HMSF  regions in the Galaxy (with L$\sim10^{5-7}$ \ls), containing clusters of OB-type stars, and are associated with compact HII regions and hot molecular cores.  
So far, we have obtained data for NGC7538~IRS1 and the W51 complex, while W3(OH), Sgr B2 (N), and G10.62 will be observed between the end of 2014 and  2015. 

The key goal of the program is to examine the kinematics and physical conditions of hot gas at very small radii from  young O-type stars, 
through the combination of subarcsecond resolutions and the high temperatures required to excite high-JK lines. 

\begin{table*}
\caption{Parameters of JVLA Observations toward NGC7538~IRS1.}             
\label{obs}      
\centering                     
\begin{tabular}{ccccccc} 
\hline\hline               
\noalign{\smallskip}
\multicolumn{1}{c}{Transition$^{a}$} & \multicolumn{1}{c}{$\nu_{\rm rest}$} & \multicolumn{1}{c}{$E_l/k^{b}$} & \multicolumn{1}{c}{Date}  & \multicolumn{1}{c}{JVLA} & \multicolumn{1}{c}{Synthesized Beam} & \multicolumn{1}{c}{RMS$^{c}$} \\ 
\multicolumn{1}{c}{(J,K)} & \multicolumn{1}{c}{(MHz)} & \multicolumn{1}{c}{(K)} & \multicolumn{1}{c}{(yyyy/mm/dd)} & \multicolumn{1}{c}{Receiver}  & \multicolumn{1}{c}{$\theta_M('') \times \theta_m(''); \ P.A.(^{\circ})$} &  \multicolumn{1}{c}{(mJy/beam)}   \\
\noalign{\smallskip}
\hline
\noalign{\bigskip}
(6,6)   & 25055.96    &  408  & 2012/05/31 & K \  \ & $0.26 \times 0.23;\ +45 $  & 1.3 \ \ \\
(7,7)   & 25715.14    &  538  & 2012/05/31 & K \ \ & $0.25 \times 0.22;\ +45 $  & 1.5 \ \ \\
(9,9)   & 27477.94    &  852  & 2012/06/21 & Ka & $0.23 \times 0.21;\ +30 $  & 2.3 \ \ \\
(10,10) & 28604.75    & 1035  & 2012/08/07 & Ka & $0.22 \times 0.20;\ +40 $  & 3.3 \ \ \\
(12,12) & 31424.94    & 1455  & 2013/12/21 & Ka & $0.33 \times 0.21;\ -76 $  & 5.5$^{d}$ \\
(13,13) & 33156.84    & 1691  & 2012/06/21 & Ka & $0.19 \times 0.17;\ +35 $  & 2.3 \ \ \\
(14,14) & 35134.28    & 1945  & 2013/12/21 & Ka & $0.30 \times 0.19;\ -76 $  & 6.0$^{d}$ \\
\noalign{\smallskip}
\multicolumn{7}{l}{Other molecular transitions}\\
\noalign{\smallskip}
\met$^{e}$ &27472.53&234&2012/06/21&Ka&$0.22 \times 0.20;\ +30 $ & 2.4 \ \ \\
CH$_3$CN$^{f}$ &36793.71&10&2012/08/07&Ka& $0.18 \times 0.15;\ +3 \ \ $ & 4.0 \ \ \\
\noalign{\smallskip}
\hline   
\end{tabular}
\tablefoot{\\
(a) Transitions include ortho-\amm~($K=3n$) and para-\amm~($K\neq3n$).  \\
(b) Energy above the ground from the JPL database.  \\ 
(c) RMS noise in a $\sim$0.4~\kms~channel. \\ 
(d)  The rms noise in the maps of transitions (12,12) and (14,14) is higher than other lines, because only 7 min on-source were spent in December 2013 (53 min were typically spent in the 2012 tracks). 
    In order to boost sensitivity, in imaging  the (12,12) and (14,14) lines we used natural weighting (as opposed to  Briggs  weighting), which resulted in  slightly larger beams. 
\\
(e)  The $J_K$= 13$_2$-13$_1$ line of CH$_3$OH was detected in the same baseband as the \amm\,(9,9) line.\\
(f)  The CH$_3$CN (2-1) line was observed in a separated baseband paired with the \amm\,(10,10) line, but was not detected.
}
\end{table*}

\subsection{NGC7538 IRS1}

This paper is the first in a series of \amm\,multilevel imaging studies in well-known HMSF regions 
and focuses on NGC 7538~IRS1. 
At a distance of 2.7 kpc \citep{2009ApJ...693..406M},   this YSO has   
 a luminosity equivalent to an O6/7 ZAMS star ($8\times10^4$ \ls) and a mass of about 30~\ms\,\citep[e.g.,][]{AkabaneKuno05}.  
Recently, a number of interferometric studies conducted with increasing angular resolution,  
at 1.3 mm with SMA (3\arcsec~beamsize, \citealt{Qiu11}),
 at 1.3 and 3.4 mm with SMA and CARMA (0\pas7~beamsize, \citealt{Zhu13}),
 and at 0.8 mm with PdBI (0\pas2~beamsize, \citealt{Beu13}) and SMA  (2\arcsec~beamsize, \citealt{Frau14}), 
detected  several typical hot-core species, showing inverse P-Cygni profiles, probing inward motion of the dense gas toward IRS1 with $\dot{M}\sim10^{-3}$~\msyr\,on scales $\gtrsim$1000 AU.  
Radio and mm studies also identified several outflows emanating from IRS1, 
along north-south or N-S \citep{Gaume95,Sandell09}, northwest-southeast or NW-SE \citep{Qiu11},  
and northeast-southwest or NE-SW \citep{Beu13}, respectively.  
The simultaneous presence of a jet/outflow and strong accretion flow toward IRS\,1, led some authors to postulate the presence of an accretion disk surrounding IRS\,1, whose evidence is mainly based on VLBI imaging of methanol masers \citep{Min98,Pes04,Surcis11}. 
Competing models  have been however proposed by different groups, to explain  positions and l.o.s. velocities of \met\,maser spots in the region, involving disk/outflow systems with different orientations. 
Based on multi-epoch VLBI observations of methanol masers, \citet{MoscaGoddi14} measured line-of-sight velocities, proper motions, and acceleration of individual maser spots in the region, building a detailed model of the 3D dynamics of molecular gas in IRS1. 
Based on this model, they proposed the existence of a multiple system of massive YSOs surrounded by accreting disks and possibly driving different outflows. 
Despite the detailed dynamical model from masers, the kinematics of thermal molecular gas as well as its physical conditions  at small radii ($<$1000 AU) from the exciting YSO(s) are largely unknown.

Using the JVLA, we imaged at a resolution of 0\pas15--0\pas3 seven \amm~lines with energy levels high above  the ground state (equivalent to 400-1950~K), from ($J,K$)=(6,6) to (14,14), in NGC7538~IRS1.
A partial dataset/analysis were presented in \citet{MoscaGoddi14}. 
  In this article we present a full report on the results and a complete analysis of the \amm\,data.

\section{Observations and Data Reduction}
\label{obser}
Observations of NH$_3$ towards NGC7538~IRS1 were conducted using the JVLA of the National
Radio Astronomy
Observatory (NRAO)\footnote{NRAO is a facility of the National Science Foundation operated under cooperative agreement by Associated Universities, Inc.} in B configuration.
By using the broadband JVLA K- and Ka-band receivers, we
observed  a total of seven metastable inversion transitions of NH$_3$: 
($J,K$)=(6,6), (7,7), (9,9), (10,10), (12,12), (13,13), (14,14)  at 1~cm (25-35~GHz).  
Transitions were observed in pairs of (independently tunable) basebands 
during 6h tracks (two targets per track) on three different dates in 2012: 
the (6,6) and (7,7) lines on May 31 at K-band, the (9,9) and (13,13) lines on June 21, and the (10,10) along with the CH$_3$CN (2--1) transition on August 7, both  at Ka-band.
The (12,12) and (14,14) transitions were observed in a 30~min track on December 21 (2013), 
as part of a test aimed at measuring the absolute position of the phase-reference source (see below). 
Table~\ref{obs} summarizes the observations and reports frequencies of all the observed transitions.  
Each baseband had 8 sub-bands with a 4~MHz bandwidth per sub-band ($\sim$40~\kms~at 30~GHz) 
and 128 channels per sub-band, providing a total coverage of 32 MHz ($\sim$320~\kms~at 30~GHz) 
and a channel separation of 31.25~kHz ($\sim$0.3~\kms~at 30~GHz). 
In December 2013, the   JVLA correlator was more flexible, allowing different settings for different sub-bands, 
therefore we used  a 16 MHz bandwidth and 1024 channels with a separation of 15.6 kHz for the sub-bands containing the lines, 
and 128 MHz bandwidth with 128 channels  for the remaining sub-bands for the continuum. 
Typical on-source integration time was 53 min in the 2012 tracks, and 7 min in the 2013 track 
(thus the different rms sensitivity reported in Table~\ref{obs}).

Each transition was observed with 
``fast switching'', where 60s scans on-target were alternated with 60s
scans of the nearby (3$^{\circ}$) QSO J2339+6010  (measured flux density 0.2--0.3~Jy, depending on frequency). 
We derived  absolute flux calibration from observations of 3C~48 ($F_{\nu}= 0.5-0.7$~Jy, depending on frequency), and bandpass calibration from observations of 3C~84 ($F_{\nu}= 24-28$~Jy, depending on frequency).

The data were edited, calibrated, and imaged in a standard fashion using the Common Astronomy Software Applications (CASA) package. 
We   fitted and subtracted continuum emission  from the spectral line data in the uv plane  
using CASA task UVCONTSUB, combining the continuum (line-free) signal from all eight sub-bands around the \amm\,line. 
We performed self-calibration of the continuum emission (typically, two phase-cycles and one amplitude-cycle), providing continuum images with an integrated flux density  and a 1$\sigma$ rms  around 400 mJy and 0.2--0.3 mJy beam$^{-1}$, from 25 to 36 GHz, respectively.  
We then applied the self-calibration solutions from the continuum to the line datasets. 
Using the CASA task CLEAN, we imaged the NGC7538 region with a cell size of 0\pas04, covering a 
8\arcsec~field around the phase-center position: $\alpha(J2000) = 23^h 13^m 45\rlap{.}^s4$,   $\delta(J2000) = +61^{\circ} 28' 10"$. 
We adopted Briggs weighting with a ROBUST parameter set to 0.5 for all transitions observed in 2012, 
resulting in a synthesized clean beam FWHM of 0\pas15--0\pas27 and an RMS noise level per  channel of $\sim$1.5--3~mJy~beam$^{-1}$ (depending on frequency). 
 In order to boost sensitivity in the images of  the (12,12) and (14,14) lines (integrated over only 7~min), we used natural weighting, which resulted in  slightly larger beams.  
  For all transitions, we smoothed the velocity resolution to 0.4 \kms. 
Table~\ref{obs} summarizes the observations.  

Since the position of the phase calibrator, J2339$+$6010, was only accurate within 0\pas15, according to the VLA Calibrator Manual\footnote{http://www.vla.nrao.edu/astro/calib/manual/csource.html}, 
we conducted a 30~min test in December 2013 to check the astrometry of our \amm\ maps, observing J2339$+$6010 and NGC7538 in the fast-switching mode using QSO J2230+6946 as a calibrator. 
We measured a positional offset of $\Delta \alpha \sim-$0\pas01, $\Delta \delta \sim$0\pas16 from the nominal position of J2339$+$6010 reported in the VLA catalog. 
All the \amm\ maps of NGC7538~IRS1 reported in this paper were shifted accordingly to the measured offset.

\begin{table}
\caption{Parameters of the \amm\, inversion lines observed toward NGC7538~IRS1.}             
\label{lines}      
\centering                        
\begin{tabular}{ccccccccclcccl} 
\hline\hline                 
\noalign{\smallskip}
\multicolumn{1}{c}{Line} &  & \multicolumn{1}{c}{F$_{\rm peak}$} &  \multicolumn{1}{c}{V$_{c}$} & \multicolumn{1}{c}{$\Delta V_{1/2}$}&\multicolumn{1}{c}{F$_{\rm int}$} &\multicolumn{1}{c}{F$_{\rm cont}$}\\ 
\multicolumn{1}{c}{(J,K)} &  &  \multicolumn{1}{c}{(Jy)}  & \multicolumn{1}{c}{(km/s)}     & \multicolumn{1}{c}{(km/s)}&  \multicolumn{1}{c}{(Jy~km/s)} &  \multicolumn{1}{c}{(Jy)}\\
\noalign{\smallskip}
\hline
\noalign{\bigskip}
 \multicolumn{6}{c}{Core Component} \\
\noalign{\smallskip}
(6,6)   &     &  -0.236 & -59.3 & 8.3 & -2.1  \ \ & { 0.34}\\
(7,7)   &     &  -0.208 & -59.4 & 7.4 & -1.6  \ \   & { 0.34} \\
(9,9)   &     &  -0.194 & -59.1 & 7.8 & -1.6  \ \   & { 0.34} \\
(10,10) &     &  -0.135 & -59.2 & 7.1 & -1.0  \ \   & { 0.37} \\
(12,12) &     &  -0.113 & -59.3 & 7.9 & -0.95 & { 0.34} \\
(13,13) &     &  -0.048 & -58.2 & 10  & -0.52 & { 0.46} \\
(14,14) &     &  -0.032 & -58.8 & 13  & -0.45 & { 0.42} \\
\noalign{\smallskip}
\met    &     &  -0.155 & -58.9 & 5.9 & -0.98  & { 0.34} \\
\noalign{\bigskip}
 \multicolumn{6}{c}{Southern Component} \\
\noalign{\smallskip}
(6,6)   &     &  -0.060 & -59.7 & 5.6 & -0.36 & { 0.07} \\
(7,7)   &     &  -0.046 & -59.8 & 4.4 & -0.21 & { 0.07} \\
(9,9)   &     &  -0.029 & -59.5 & 4.4 & -0.14 & { 0.07} \\
(10,10) &     &  -0.013 & -59.4 & 4.3 & -0.06 & { 0.07} \\
\noalign{\smallskip}
\met    &     &  -0.044 & -59.6 & 4.6 & -0.22 & { 0.07} \\
\noalign{\smallskip}
\hline   
\end{tabular}
\tablefoot{
The peak fluxes (F$_{\rm peak}$, col. 2), the central velocities (V$_{c}$; col. 3), the FWHM line-width ($\Delta V_{1/2}$; col. 4), and the velocity-integrated flux (F$_{\rm int}$; col. 5) are estimated from single-Gaussian fits to the spectral profiles shown in Figures~\ref{spectra-1pan} and \ref{spectra-ch3oh}.
The core shows actually a stronger component around -59~\kms\,and a weaker component 
around -56~\kms, the latter becomes more comparable with the stronger component at higher excitation energies. 
This partly explains the different velocity-centroids and line-widths at different excitations. 
 The table reports also the continuum emission flux density at the corresponding frequency of the inversion line (F$_{\rm cont}$; col. 6). 
 }
\label{nh3_lines}
\end{table}

\section{Results}
\label{res}
For the first time we have mapped the hot \amm~gas from metastable transitions (6,6) up to (14,14) with 0\pas3-0\pas15 resolution towards NGC7538~IRS1. 
These transitions are observed in absorption against the strong HC-HII region associated with  NGC7538~IRS1. 
In particular, we observed three ortho (6,6; 9,9; 12,12) and four para (7,7; 10,10; 13,13; 14,14) transitions, with lower-state energy from approximately 400 K to 1950 K\footnote{It is worth noting that the (13,13) and (14,14) lines are at 1691 K and 1945 K above the ground, and therefore provide the lines with the highest energy above the ground state ever imaged with an interferometer so far (these lines have been detected with the 100-m telescope in Orion BN/KL and Sgr-B2 by \citealt{Wilson93} and \citealt{Huettemeister95}, respectively).}.  
This enables us to probe various temperature components of the hottest molecular gas in the region.
Remarkably, for the lower-excitation lines, from (6,6) to (9,9), we even detect pairs of hyperfine quadrupole satellites,
which enabled direct estimation of the optical depth (see \S~\ref{tau}).

Besides \amm, we detected also  the $J_K$= 13$_2$-13$_1$ line of CH$_3$OH (with a rest frequency of 27.47253 GHz).  The CH$_3$CN (2-1) line included in our bandwidth (with a rest frequency of 36.79371 GHz) was not detected. 

The parameters of all the observed transitions are reported in Tables~\ref{obs} and \ref{nh3_lines}.  

For each transition, we derived spectral profiles (Sect.~\ref{spec}), maps of total intensity  (Sect.~\ref{intensity}) and velocity field (Sect.~\ref{velocity}), position-velocity diagrams (Sect.~\ref{pv}), and velocity-channel centroid maps  (Sect.~\ref{spots}). 

\begin{table*}
\caption{Parameters of the hyperfine components of \amm\, inversion lines observed toward NGC7538~IRS1.}             
\label{hf_lines}      
\centering                        
\begin{tabular}{cccccccccc} 
\hline\hline                 
\noalign{\smallskip}
\multicolumn{1}{c}{Line} & \multicolumn{2}{c}{$\Delta \nu_{HF}$}& \multicolumn{2}{c}{$\Delta V_{HF}$} &   \multicolumn{1}{c}{$a_{ms}$}&  \multicolumn{1}{c}{F$_{\rm peak}$} &  \multicolumn{1}{c}{$\Delta V_{1/2}$}&\multicolumn{1}{c}{F$_{\rm int}$} 
& Opacity 
\\ 
\multicolumn{1}{c}{(J,K)} & \multicolumn{2}{c}{(MHz)}  & \multicolumn{2}{c}{(\kms)}  & & \multicolumn{1}{c}{(Jy)} & \multicolumn{1}{c}{(km/s)}      &  \multicolumn{1}{c}{(Jy~km/s)} 
& ($\tau$)
\\
& Inner & Outer & Inner & Outer&&&&&\\
\noalign{\smallskip}
\hline
\noalign{\bigskip}
 \multicolumn{6}{c}{Core Component} \\
\noalign{\smallskip}
(6,6)   &  $\pm$2.24 & $\pm$2.62   &$\pm$26.9 & $\pm$31.4  & 0.0081 & -0.060 & 7.6 & -0.4  \ \  & 26 \\
(7,7)   &  $\pm$2.34 & $\pm$2.68   &$\pm$27.3 & $\pm$31.2  & 0.0060 & -0.022 & 6.6 & -0.2  \ \  & 22 \\
(9,9)   &  $\pm$2.48 & $\pm$2.75   &$\pm$27.0 & $\pm$30.1  & 0.0037 & -0.010 & 6.9 & -0.06& 10 \\
\noalign{\bigskip}
 \multicolumn{6}{c}{Southern Component} \\
\noalign{\smallskip}
(6,6)   &  $\pm$2.24 & $\pm$2.62   &$\pm$26.9 & $\pm$31.4  & 0.0081 & -0.007 & 4.5 & -0.034 & 12 \\
(7,7)   &  $\pm$2.34 & $\pm$2.68   &$\pm$27.3 & $\pm$31.2  & 0.0060 & -0.002 & 5.2 & -0.012 & 10 \\
\noalign{\smallskip}
\hline   
\end{tabular}
\tablefoot{
  The  frequency separations ($\Delta \nu_{HF}$, cols.~2 and 3) of the four satellite components can be calculated 
using quantum mechanics formalism for a symmetric molecular rotor.  
The corresponding velocity separations ($\Delta V_{HF}$) are also reported in cols.~4 and 5. 
$a_{ms}$ (col.~6) is the theoretical ratio of the satellite line  to the main line strengths. 
F$_{\rm peak}$, F$_{\rm int}$, and $\Delta V_{1/2}$ were fitted simultaneously for all the  hyperfine quadrupole satellite components assuming a Gaussian shape for each of the hyperfine lines.
The values reported in the table are average values of the four Gaussians fitted to the HFS satellites. 
$\tau$ (col. 10) is the line opacity estimated numerically with Equation~\ref{tau_eq}. 
}
\label{nh3_hf}
\end{table*}

\begin{figure}
\centering
\includegraphics[width=0.42\textwidth]{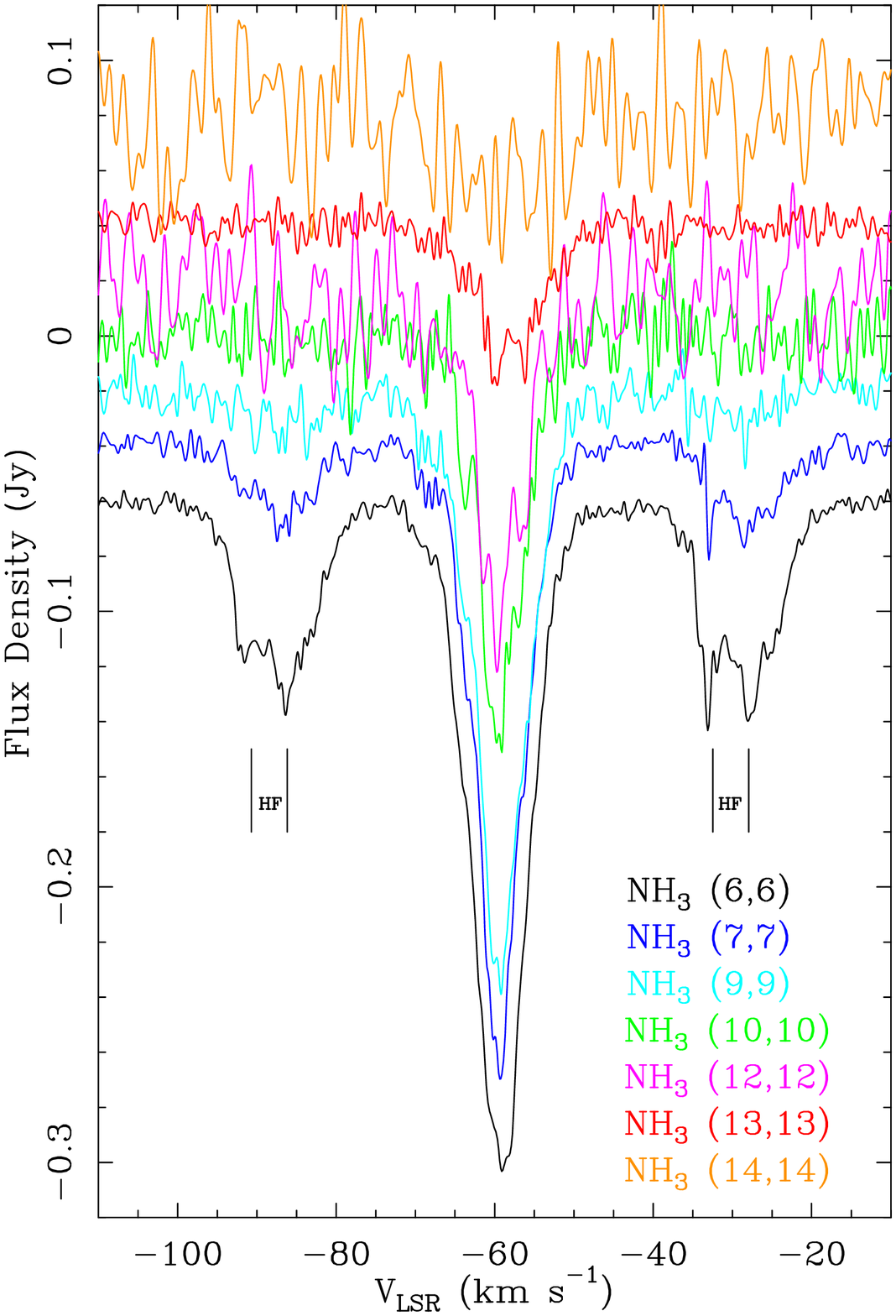}
\includegraphics[width=0.42\textwidth]{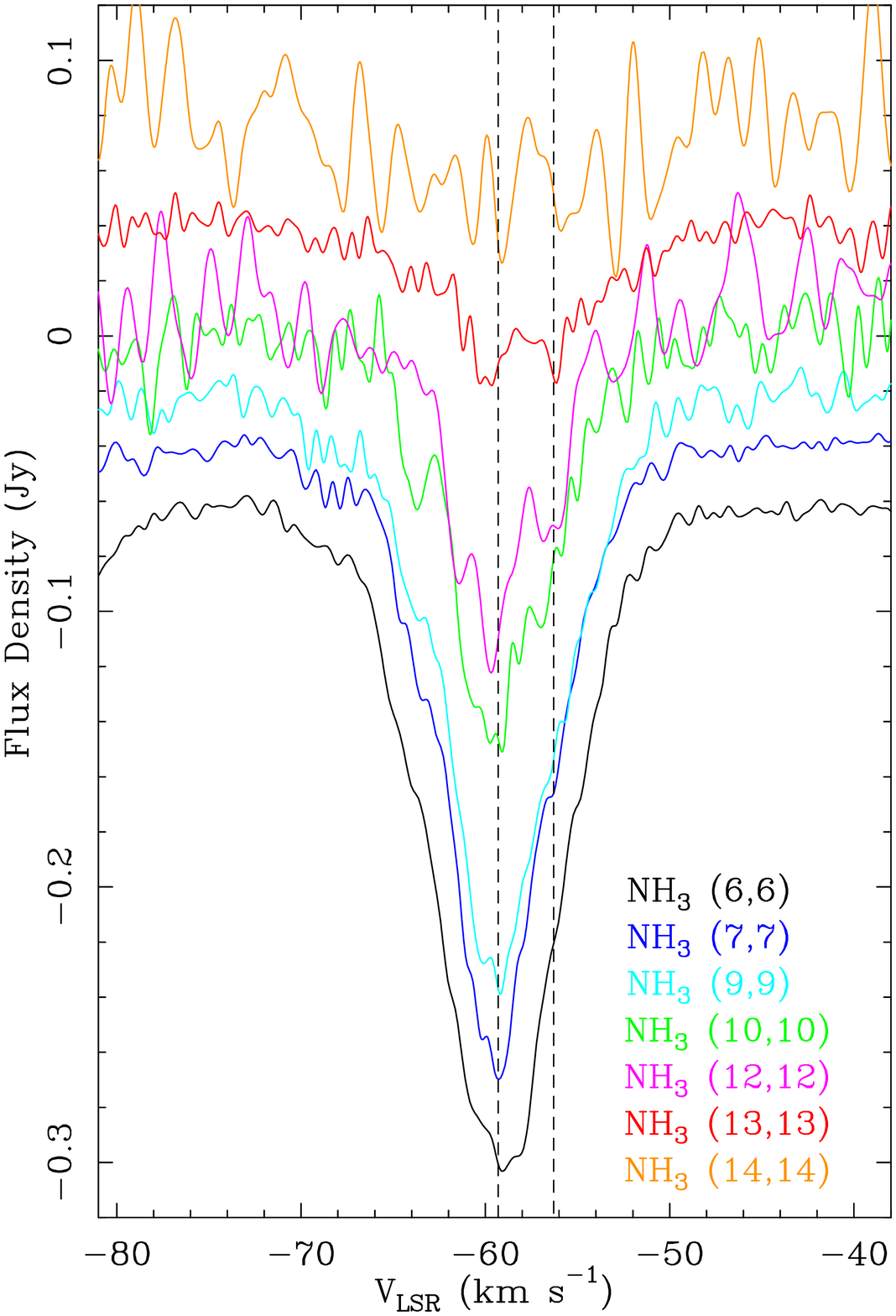}
\caption{Spectra of various NH$_3$ inversion transitions integrated over the compact core only of NGC7538~IRS1.  
Spectral profiles for transitions (6,6), (7,7), (9,9), (10,10), (12, 12), (13,13), and (14,14) are shown in the same panel, 
where an offset in flux density of 0.02 Jy is applied to transitions adjacent in energy, to better display individual profiles.  
The lower state energy levels of transitions shown here are $\sim 408-1947$\,K (see Table~\ref{obs}).
The velocity resolution is 0.4~\kms. 
({\it Upper panel}) The hyperfine satellite lines, separated by $\sim \pm24-31$ \kms\,(see Table~\ref{nh3_hf}), are clearly detected for the (6,6), (7,7), and (9,9) lines.
({\it Lower panel}) A narrower velocity range is displayed here, in order to better evidence the line profiles 
of the main hyperfine component of each inversion transition. 
The vertical dashed lines indicate the two main velocity components at --59.3~\kms\,and 56.4~\kms\,(see text).  
}
\label{spectra-1pan}
\end{figure}
\begin{figure}
\centering
\includegraphics[width=0.45\textwidth]{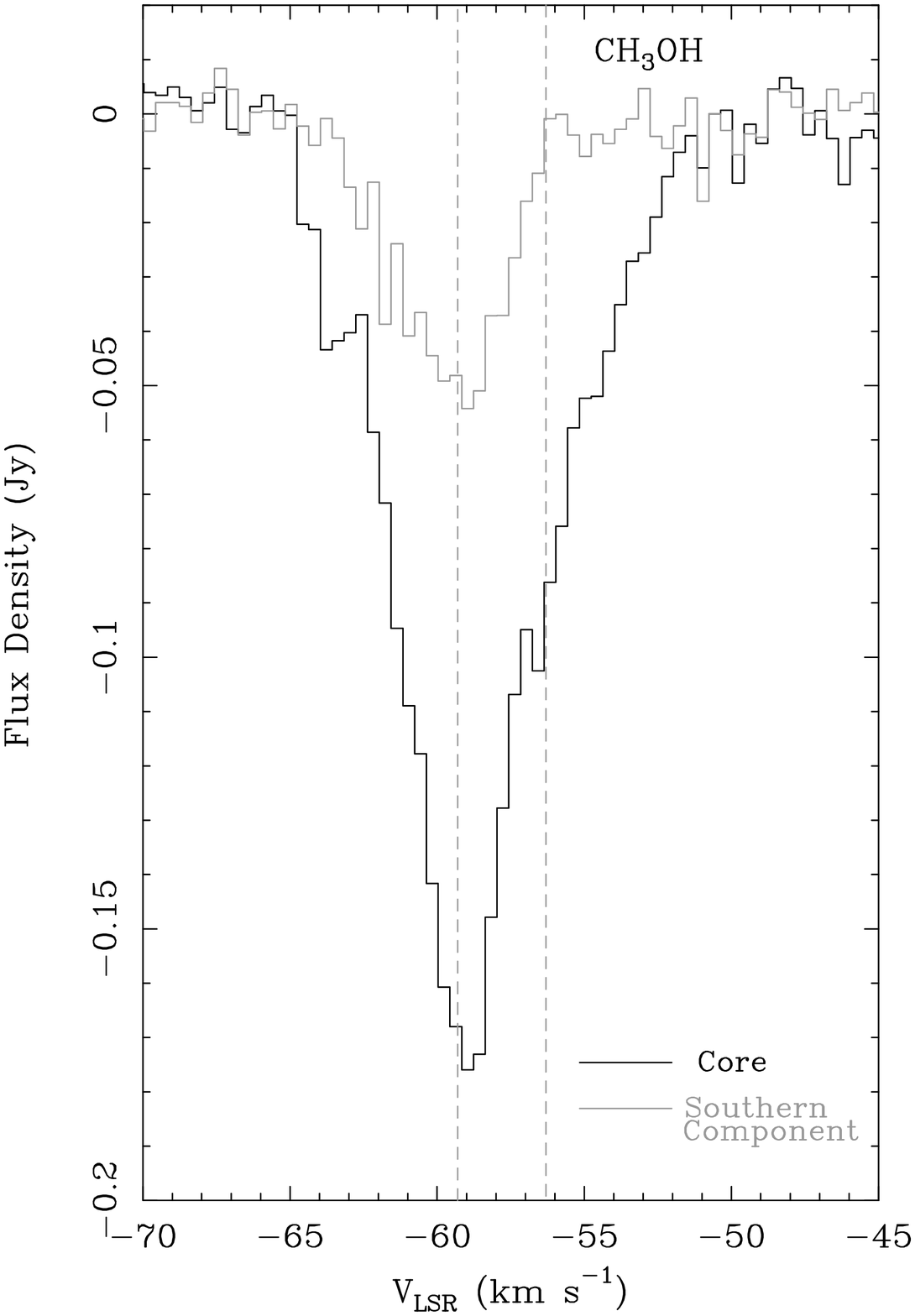}
\caption{Spectral profile of the $J_K$= 13$_2$-13$_1$ line of CH$_3$OH ($\nu_{rest}=27.473$ GHz), 
integrated over the core component  (black line) 
and the southern spherical component (grey line).
The vertical dashed lines indicate velocities of --59.3~\kms\,and 56.4~\kms, 
corresponding to the two main velocity components in the core (see text). 
The velocity resolution is 0.4~\kms. 
}
\label{spectra-ch3oh}
\end{figure}

\subsection{Spectral Profiles}
\label{spec}
\begin{figure*}
\centering
\includegraphics[angle=-90,width=\textwidth]{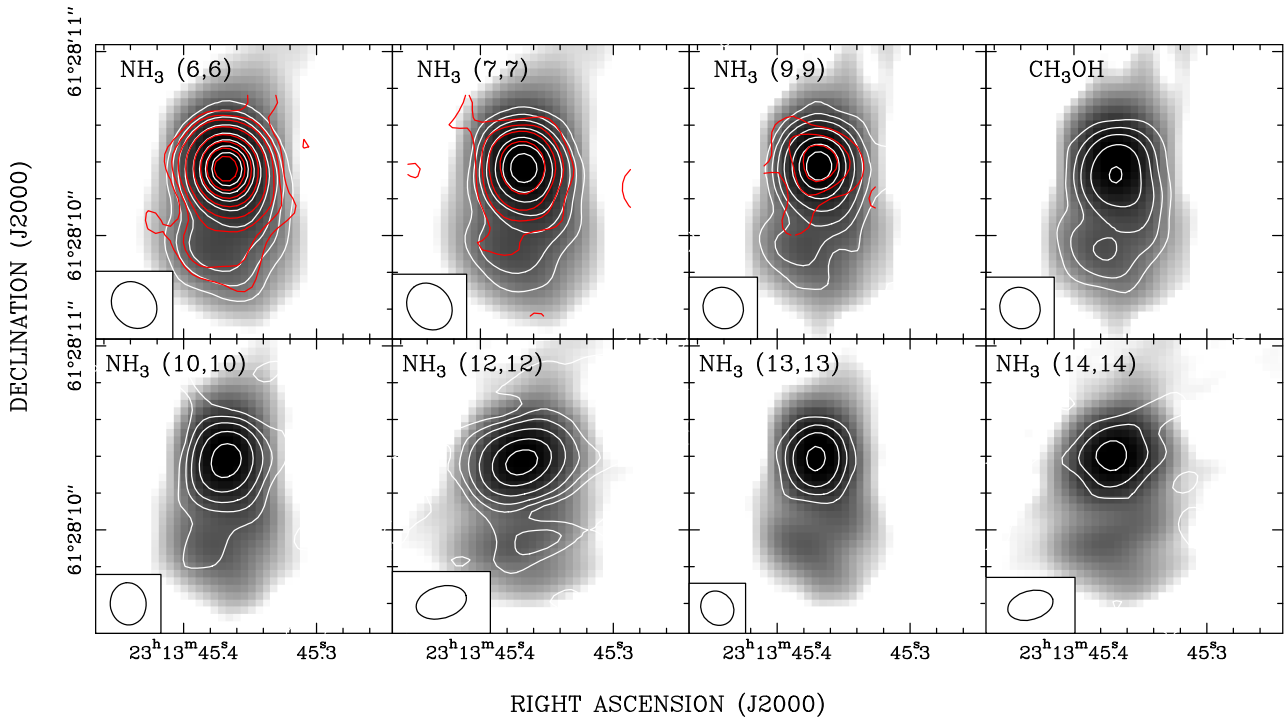}
\caption{Total intensity images of seven inversion transitions of  NH$_3$ as well as the \met\,line, as observed toward NGC7538~IRS1 with the JVLA B-array. Contours of the \amm\,line absorption are overplotted on the 
continuum emission (black image) at the frequency corresponding to each line. 
 The contours represent factors 1, 3, 6, 12,....(afterwards, they are spaced by 10) of --20~mJy~beam$^{-1}$, for all transitions.  
The images were integrated over the velocity range 
--71~\kms~to --47~\kms, covering only the main hyperfine component.  
  The line to continuum ratio decreases from 0.7 for the (6,6) transition to  0.1 for the (13,13) line (Table~\ref{nh3_lines}) . 
For the (6,6), (7,7), and (9,9) lines we overplotted also the absorption from one pair of hyperfine components (red contours).  
The velocity resolution was smoothed to 0.4~\kms\,for all transitions. 
The images were constructed with a 0\pas04 pixel for all transitions. 
No flux cutoff was applied.  
The synthesized beams (0\pas17--0\pas33)  are shown in the lower left corner of each panel (see Table~\ref{obs}). 
The lower state energy levels of transitions shown here are 408-1947\,K.
}
\label{mom0}
\end{figure*}

For each transition, we produced spectra by mapping each spectral channel and summing the flux density in each channel map (Figures~\ref{spectra-1pan}, \ref{spectra-ch3oh}). 
Since we resolve IRS1 in the central core and a southern component (see \S~\ref{intensity} and Fig.~\ref{mom0}), 
we extracted spectral profiles towards both components of the radio continuum.  
In the core, multiple transitions of \amm, from (6,6) to (12,12), show similar central velocities (V$_c=$--59.1  to --59.4~\kms), 
and FWHM line-widths ($\Delta V_{1/2}$=7.1--8.3~\kms) of the main hyperfine component,  
as determined from single-Gaussian fits (Table~\ref{nh3_lines}). 
We assume --59.3~\kms\,as the systemic velocity of the core. 
The most highly excited lines, the (13,13) and (14,14) lines, display larger line-widths (10 \kms\,and 13 \kms) 
and more redshifted central velocities (--58.2 \kms\,and --58.8 \kms). 
This may reflect lower signal-to-noise ratio in the Gaussian fitting. 
Alternatively, these differences may reflect dynamical properties of the hottest molecular gas,   
 expected to be closer to the  central YSO and therefore to move faster.  
A third possibility is the presence of multiple velocity components. 
Figure~\ref{spectra-1pan} (lower panel) shows in more detail the profiles of the main hyperfine component 
from different \amm\,transitions towards the core. 
These profiles show indeed two main velocity components, the dominant one, around --59.3~\kms\,(the systemic velocity), 
and a weaker one, around --56.4~\kms, which appears to become more comparable to the main component 
 with higher excitation energy (for the 13,13 line they show similar strengths).  
This explains the progressive increase of the central velocity (and line-width) with the excitation energy, 
when only one Gaussian is employed for the fit of the spectral profiles. 
The two velocity components are also evident in the \met\,spectral profile towards the core 
(see Figure~\ref{spectra-ch3oh}).  
Since in the case of \met\,the relative intensity of the weaker component is higher 
with respect to the lower excitation \amm\,lines, 
single-Gaussian fitting provides a more redshifted central velocity  (see Table~\ref{nh3_lines}). 
Besides the presence of two components, the profiles are quite symmetric and can be reasonably fitted by a Gaussian profile\footnote{
We also attempted a two-component fitting, which reported smaller line-widths, $\sim$3~\kms\,and $\sim$5--6~\kms\,for the dominant and weaker component, respectively. 
However, since the \amm\,spectral profiles show, besides the main component,  a hyperfine structure with two extra pairs of weak satellites, 
 the signal-to-noise ratio in the latter did not warrant two-component fitting, 
 resulting in very unreliable estimates of  physical parameters (see \S~\ref{phys}), 
such as the optical depth, calculated from the satellite to the main line intensity ratio 
(this implies that a small error in the line intensity of the satellite results in a large change in the opacity).   
Therefore, in the analysis presented in this paper we adopted one component fitting.}. 

Our results are consistent with previous spectroscopic studies. 
Based on CARMA and SMA observations, \citet{Zhu13} report a similar value of --59.5~\kms\, for the systemic velocity,  employing low-excitation lines at 1.3~mm ($E_{\mathrm {up}}\sim70-500$ K), 
but also find an average peak velocity of --58.6 \kms, using more highly excited (and more optically thin) lines at 0.86 mm. 
Interestingly, the spectral profile of the $^{13}$CO (2-1) line shows multiple velocity components in absorption, 
with the strongest ones at --59~\kms\,and --57~\kms, consistent with our findings. 
Similar velocity components are observed in absorption in many other molecules imaged with the PdBI at 1.36~mm 
(e.g., OCS 18-17, HC$_3$N 24-23, NH$_2$CHO 10-9, CH$_3$OH 4-3; \citealt{Beu12}, their Fig.~7). 
Finally, \citet{Knez09} analyzed high-resolution mid-IR spectra with absorption lines from several molecules 
(C$_2$H$_2$, CH$_3$, CH$_4$, NH$_3$, HCN, HNCO, CS) 
and found two Doppler shift components at --59.7~\kms\,and --56.4~\kms, very close to the values observed in \amm.  
These velocities also correspond to the velocity-centroids of two methanol maser clusters separated by 0\pas2 as imaged with VLBI \citep{MoscaGoddi14}. 

Previous single-dish observations of the (7,7) inversion line with the Effelsberg 100-m telescope, 
provided a line flux density of $\sim$260 mJy (assuming an equivalence relation K=0.86~Jy; \citealt{Mauersberger86}), so we recovered more than 80\% of the single-dish flux density. 
Therefore, we can safely assume that most of the flux density for transitions from (6,6) to (14,14) is recovered by the  JVLA B-array. 
This is not surprising considering that these transitions arise from levels $>$400~K above the ground and are seen in absorption against the compact HII region.

Owing to interaction with the quadrupole moment of the nitrogen nucleus, 
each \amm\,inversion line is split into five components, 
a ``main component'' and four symmetrically spaced ``satellites'', 
which make up the quadrupole hyperfine structure (HFS). 
The relative intensities and frequency separations of the four satellite components can be calculated 
using quantum mechanics formalism for a symmetric molecular rotor (see Table~\ref{nh3_hf}).  
The four satellite components are predicted to have nearly equal intensities, at least in LTE. 
In NGC7538~IRS1, we found prominent hyperfine  satellite lines in the lower-excitation doublets, 
from (6, 6) to (9, 9), spaced $\sim$27-31 \kms\, from the main components  
(Figure~\ref{spectra-1pan}). 
We used five-component Gaussian models to fit the HFS in each of those inversion transitions, 
where we fixed their velocity separations according to the calculated values (Table~\ref{nh3_hf}). 
Fitting of HFS  was done in CLASS, part of the GILDAS software\footnote{see http://www.iram.fr/IRAMFR/GILDAS}. 
The (calculated and measured) line parameters of the hyperfine components are given in Table~\ref{nh3_hf}. 
Despite the fact that the satellites are well resolved from the main lines, 
 each pair of satellites has components with a frequency difference of only $\sim$0.3-0.4 MHz or 3-5 \kms, 
therefore they are blended together. 
For this reason, we refrain from using these satellite components to study the gas kinematics.

\subsection{Total Intensity Maps}
\label{intensity}
We produced total intensity images of the main component of various NH$_3$ inversion
transitions, integrated over their line-widths. 
In  Fig.~\ref{mom0},  the integrated intensity images of \amm\,(contours) are overlaid on the continuum emission at the corresponding frequency of the inversion line (black images).

The \amm\, absorption follows closely the continuum emission, as expected. 
In particular, for lower excitation transitions, (6,6; 7,7; 9,9), 
the absorption is extended N-S across $\sim1^{\prime \prime }\times 0.7^{\prime \prime }$ ($1900 \times 2700$~AU), 
and reveals two main condensations of hot molecular gas  associated with different peaks of the radio continuum emission, 
the "core" and the "southern spherical" components identified by \citet{Gaume95} at 1.3 cm.   
The highest excitation \amm~lines, (12,12) and upwards, originate exclusively from the  core of the radio continuum 
and probe the hottest gas associated with the hot molecular core in NGC7538~IRS1.  
 The southern spherical component has the weakest integrated absorption 
 and it is not detected in the highest-$JK$ transitions, 
 indicating lower temperatures and/or densities (see \S~\ref{phys}).   
 A detection of the satellite hyperfine components in the (6,6) and (7,7) lines indicates that 
 also this southern component is optically thick, although the estimated opacity values 
 are lower than those for the central core (Table~\ref{nh3_hf}, \S~\ref{tau}). 
 
We fit an elliptical Gaussian to the strong core component. 
In all transitions, the component is fairly compact and we obtain an average deconvolved FWHM size of 0\pas24$\times$0\pas14. 
At the source distance of 2.7 kpc, this angular size corresponds to a linear size of $\sim$500 AU.

\begin{figure*}
\centering
\includegraphics[angle=-90,width=\textwidth]{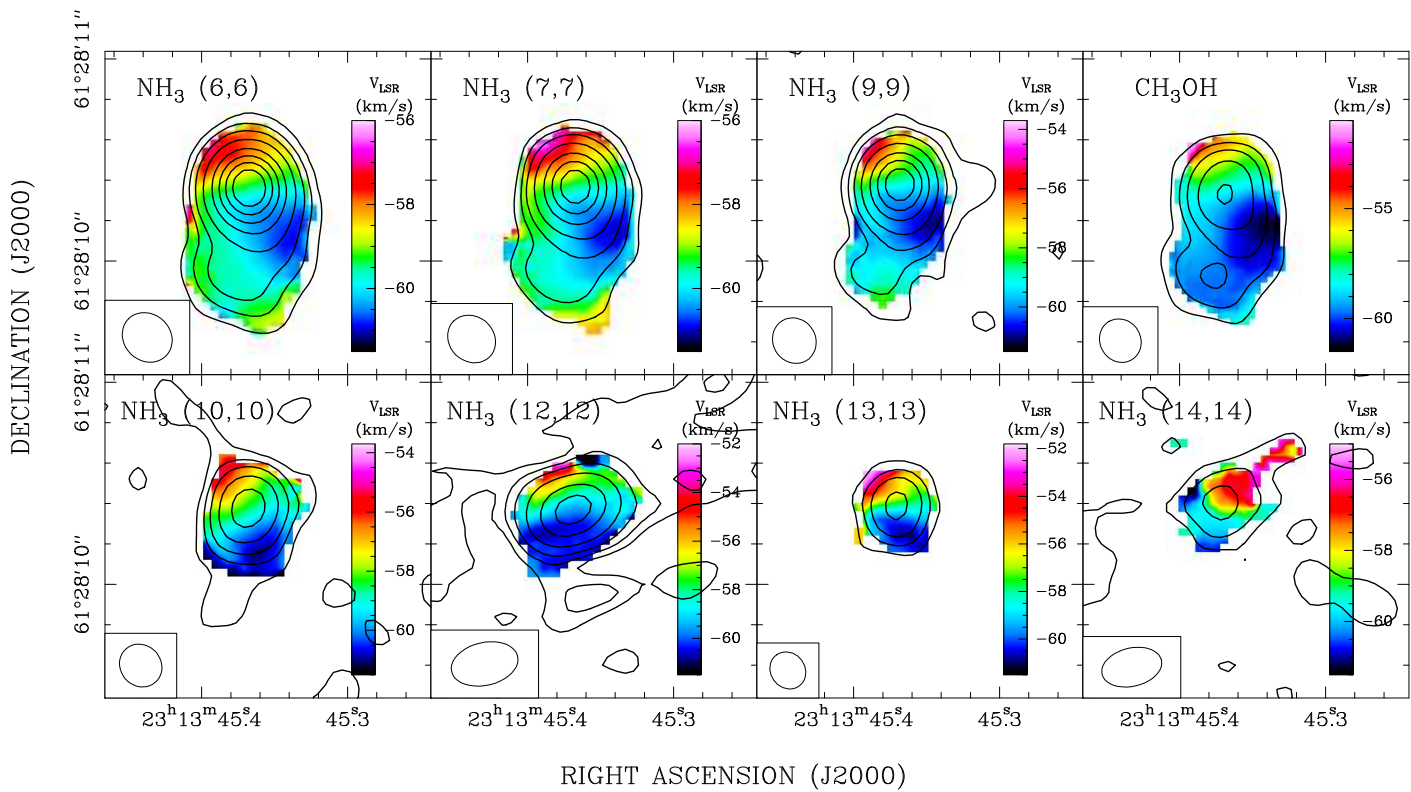}
\caption{Velocity fields of  seven inversion transitions of  NH$_3$ as well as the \met\,line, as measured toward NGC7538~IRS1 with the JVLA B-array.  
The 1$^{st}$ moment maps ({\it images}) are overlaid on the total intensity 0$^{th}$ moment maps ({\it contours}). 
Colors indicate V$_{LSR}$ in \kms. The images were constructed with a 0\pas04 pixel for all transitions. 
 The contours represent factors 1, 3, 6, 12,....(afterwards, they are spaced by 10) of --20~mJy~beam$^{-1}$, for all transitions.   
A flux cutoff of --7 to --11~mJy~beam$^{-1}$ ($\sim3-4\sigma$) was used for various transitions, 
except the (12,12) and (14,14) transitions, for which the cutoff was set to 20~mJy~beam$^{-1}$. 
Note that different velocity scales are used in different panels; in particular, color scales are expanded towards higher velocities with increasing quantum number to clearly show the redshifted absorption, e.g., \amm\,(13,13). 
All transitions show a velocity gradient NE-SW from NGC7538~IRS1, at approximately 30--40\degree\,(depending on transition), with the exception of the (14,14) line.
The synthesized beams (0\pas17--0\pas33)  are shown in the lower left corner of each panel (see Table~\ref{obs}). 
}
\label{mom1}
\end{figure*}

\subsection{Velocity Field Maps}
\label{velocity}
We also derived the kinematics of the molecular gas associated with IRS1. 
In Figure~\ref{mom1}, we show the intensity-weighted  velocity fields (or first moment maps), 
for all \amm~transitions as well as the \met\,line. 
In the figure, the velocity (in colors) is superimposed on the total intensity map (in contours), for each transition. 

Remarkably, the \amm~absorption towards the core shows an apparent velocity gradient in each line,  
with redshifted absorption towards NE and blueshifted absorption towards SW  
with respect to the hot core center (with the exception of the 14,14 line). 
The velocity gradient is at a position angle [P.A.] $\sim$30--40\degree, 
and has a magnitude $\Delta V\sim 5 \rightarrow 9$ \kms, going from the lower-excitation to the higher excitation lines. 
A  velocity gradient at the same position and with similar orientation and amplitude was measured by \citet{Zhu13} 
from SMA images of high-density tracers such as the OCS (19-18), CH$_3$CN (12-11), and $^{13}$CO (2-1) lines with 0\pas7 angular resolution, 
and  by \citet{Beu13} in PdBI images of highly excited lines such as the \ HCN (4--3) v$_2$ = 1 \ and \ the  \met\,(15$_{1,14}$--15$_{0,15}$) transitions. 
The \amm\,(14,14) line is an exception, since it shows an apparent velocity gradient  along the SE-NW direction. 

Interestingly, the \amm\,velocity gradient becomes apparently steeper with the excitation energy (Figure~\ref{mom1}), 
as one would expect if gas closer to the star moves faster. 
The most natural explanation for the observed velocity gradient would be a rotating disk. 
However, as discussed in \S~\ref{spec}, multiple velocity components are present, 
which may mimic the presence of a velocity gradient
or alter its orientation/amplitude, if a gradient is truly present. 
We will discuss these points in \S~\ref{spots} and~\ref{disk}.

\subsection{Position-Velocity Maps}
\label{pv}
\begin{figure}
\centering
\includegraphics[width=0.5\textwidth]{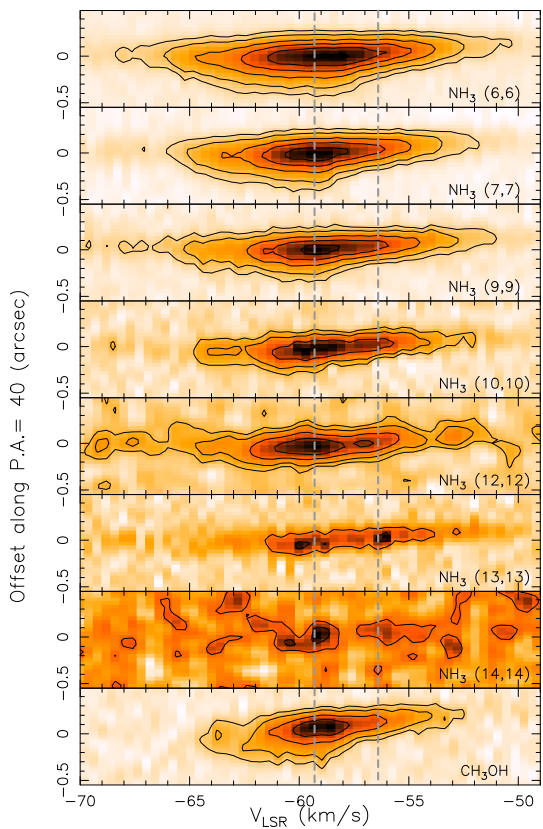}
\caption{Position-velocity diagrams of the (J,K) = (6,6), (7,7). (9,9), (10,10), (12,12), (13,13), (14,14) lines as well as the CH$_3$OH line observed with the JVLA B-array towards NGC7538~IRS1. 
The cut is taken at the peak of the \amm\,core with a P.A. = 40\degree\,for the \amm\,(6,6), (7,7). (9,9), and CH$_3$OH lines, 
and P.A. = 35\degree\,for the \amm\,(10,10), (12,12), (13,13), and (14,14). 
The contours are drawn at steps of --0.008  Jy beam$^{-1}$ starting from -0.004 Jy beam$^{-1}$.  The images were constructed with a 0\pas04 pixel for all transitions. 
Only the  main hyperfine line is shown for each transition. 
The vertical dashed lines indicate the velocities of --59.3 and 56.4 \kms. 
The P.A. is measured from North to East.
}
\label{pv_40}
\end{figure}

\begin{figure}
\centering
\includegraphics[angle=-90,width=0.5\textwidth]{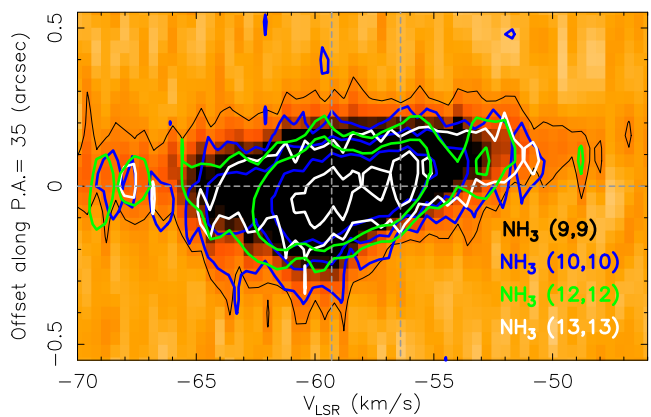}
\caption{Overlay of position-velocity diagrams of the (J,K) = (9,9), (10,10), (12,12), (13,13) lines observed with the JVLA B-array towards NGC7538~IRS1. The cut is taken at the peak of the \amm\,core with a P.A. = 35\degree, except the (9,9) line (taken at P.A. = 40\degree). 
The image shows the pv-diagram for (9,9) line. 
The contours are -0.004, --0.012, --0.024 Jy beam$^{-1}$ for all transitions, 
except the (9,9) transition that shows only the contour at -0.004 Jy beam$^{-1}$  and 
the (12,12) transition that shows only the contours at --0.012 and --0.024 Jy beam$^{-1}$, respectively.   
The vertical dashed grey lines indicate the velocities of --59.3 and 56.4 \kms. 
The P.A. is measured from North to East.
}
\label{pv_991013}
\end{figure}

 To investigate the nature of the motion in NGC7538~IRS1, we present position-velocity (pv) plots of the \amm\, inversion lines 
 in Figures~\ref{pv_40} and \ref{pv_991013}. 
The cuts  are taken at the peak of the \amm\,core along the direction of the main velocity gradients observed in the velocity field maps, P.A. = 35--40\degree.    
 The velocity gradients noticed in the  velocity maps are evident in the pv-diagrams as well.  
It is instructive to compare different lines. 
While the lower-excitation lines,  (6,6), (7,7) and even the (9,9), are dominated by a strong central component around --59~\kms, 
with increasing excitation energy several components become evident, 
with the most prominent, besides the one at  --59~\kms, at --56,--57~\kms\,(e.g., Fig.~\ref{pv_991013}). 
This is consistent with the spectral profiles and  suggests the presence of multiple individual objects moving at slightly different l.o.s. velocities (see discussion in \S~\ref{multiple}). 

Figure~\ref{pv_991013} shows an overlay of pv-diagrams for the most highly-excited lines: (J,K) = (9,9), (10,10), (12,12), and (13,13)\footnote{We excluded the (14,14) lines because of  a too low signal-to-noise ratio.}. 
While the presence of a velocity gradient is clear from this plot, there is no evident steepening of the gradient with excitation energy.  
In fact, different transitions seem  to occupy the same locus in  pv-space. 
It is worth noting that while the structure in the negative velocity - negative offset quadrant could in principle be consistent with Keplerian rotation, 
the asymmetry in the opposite quadrant indicates a more complex (non-Keplerian) profile. 

%

\subsection{Velocity-channel centroid maps}
\label{spots}

Although the core of NGC7538~IRS1 is not resolved in our \amm\,images (with a  resolution of 0\pas2 or 500 AU),  
when observed with the VLA in A-configuration (with a  resolution of 0\pas08 or 200 AU), 
the 1.3~cm continuum is  resolved in two components, 0\pas2 apart   \citep{MoscaGoddi14}. 
In the following of this Section, we refer to these two components as "northern" and "southern" components\footnote{The latter should be distinguished from the southern spherical component, first identified by \citet{Gaume95}, separated by 0\pas4 or 1000 AU from the core, from which is resolved in our \amm\,images with 0\pas2 resolution.}. 

In order to compare more in detail the structure and velocity distribution of the \amm\,gas
with the higher resolution continuum emission, 
we made velocity-centroid maps by performing two-dimensional Gaussian fitting to the absorption peak in individual velocity channels maps.  
The positional uncertainty of the absorption centroids is proportional to the synthesized beam size and is inversely proportional to the signal-to-noise ratio of the channel maps. 
With typical signal-to-noise ratios of 10--100, the positional accuracy is estimated to be of order of a few mas 
(the formal errors in the Gaussian fitting, $1\sigma=$0\pas001--0\pas02, are consistent with this expectation).
Therefore, Gaussian-fitting the positions of the absorption peaks in individual spectral channels, 
enables us to ``super-resolve'' the \amm\,gas distribution and to better compare it with the higher resolution continuum emission. 

\begin{figure*}
\includegraphics[angle=-90.0,width=\textwidth]{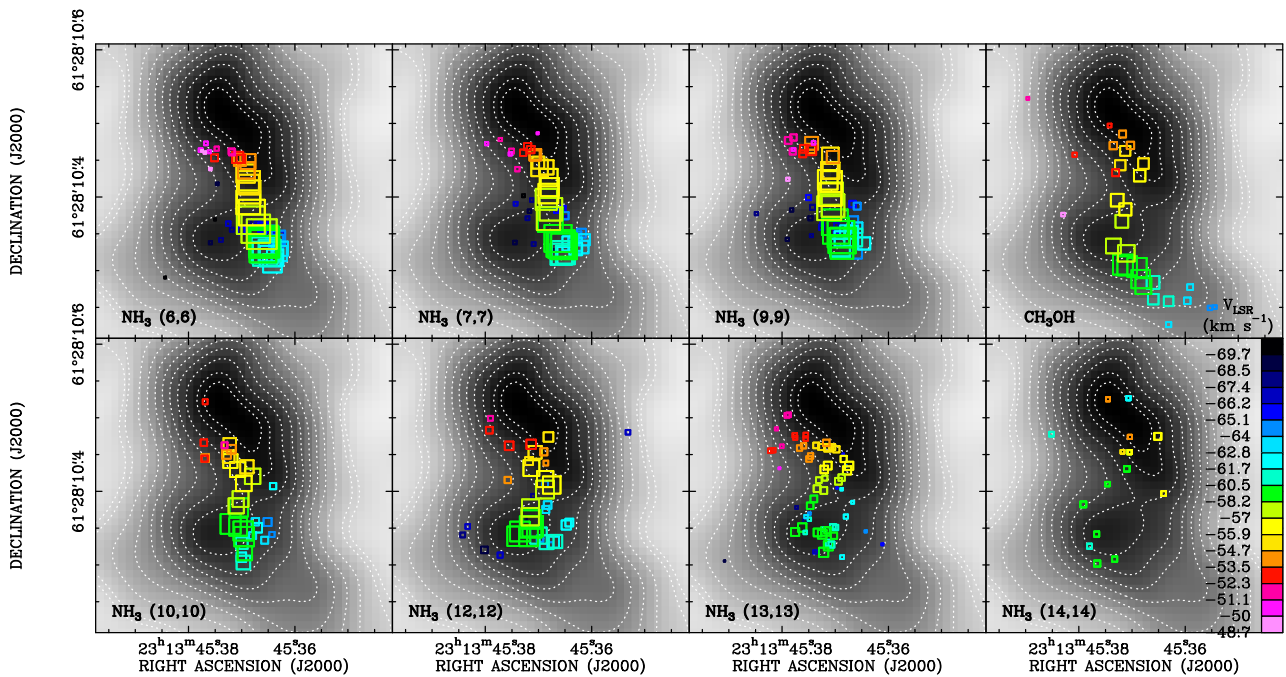}
\caption{ Centroids of \amm\,absorption fitted in individual velocity channels ({\it open squares}), 
overlaid on the  1.3~cm continuum map  ({\it black image and white contours}) imaged with the VLA in A-configuration \citep{MoscaGoddi14}. 
 {\em Color} denotes  line-of-sight velocity (color scale on the right-hand side). The sizes of squares   scale linearly with the flux density of \amm\,absorption. 
 The relative alignment between \amm\ and 1.3 cm continuum is accurate to $\sim$30~mas. 
 Note that with a linear resolution of 200 AU, the compact core is resolved in two (northern and southern) components, 
 that are unresolved in our VLA B-array images (Figures~\ref{mom0} and \ref{mom1}). 
 The  southernmost component, separated by $\sim$0\pas4 from the core, is not shown here.  
 Note that the \amm\ absorption distributes between the two peaks of the radio continuum, 
 shows a velocity gradient roughly N--S, and is strongest towards the southern component of the radio continuum.}
\label{spots_map}
\end{figure*}

Figure~\ref{spots_map} shows  positions,  velocities, and the intensity of the absorption peaks in different velocity channels 
for each individual inversion line, overlaid on the 1.3~cm continuum imaged with the VLA in A-configuration\footnote{  
In Figure~\ref{spots_map} we do not show the more southern component, separated by $\sim$0\pas4 from the core, shown in our VLA-B array images (see Figures~\ref{mom0} and \ref{mom1}).}.  

Figure~\ref{spots_map} illustrates four points: 
1)~The absorption of all the \amm\ lines is mainly distributed N-S, in between the northern and southern components of the core. 
2)~The absorption occurs at well separated velocities, being redshifted (in the range --56 to --50~\kms) in the northern component, 
and blueshifted (in the range --65 to --57~\kms) in the southern component. 
3)~The strongest absorption occurs at blueshifted velocities (peak around --59~\kms) towards the southern component.
4)~The absorption in between the two continuum peaks shows that the gas  velocity steadily increases, 
going from the southern to the northern component.

There are remarkable differences between the velocity-centroid maps of Fig.~\ref{spots_map} and the $1^{}st$ moment maps in Fig.~\ref{mom1}.   
First, in the intensity-weighted velocity maps the largest velocity gradients (with the most redshifted absorption) are observed in the highest excited lines, 
going from (--61, --56~\kms) for the (6,6) line to  (--61, --52~\kms) for the (13,13) line, 
suggesting a potential steepening of the velocity gradient with the excitation energy (see \S~\ref{velocity} and Fig.~\ref{mom1}). 
The velocity-centroid maps show instead a similar spatial distribution and 
velocity extent of the \amm\,absorption peaks among the various inversion lines, 
indicating a similar velocity gradient for lines at different excitation. 
Second, in the  $1^{}st$ moment maps the higher velocities are suppressed because of weaker signal, 
while the channel peak maps reveal the full extent of the gradient, 
 from \ $\approx$ $-$65~\kms (at the center of the southern peak) to \ $\approx$ $-$50~\kms (close to the northern peak).
Finally, the  orientation of the  velocity gradient seems to change from NE-SW in the 1$^{st}$ moment maps (Figure~\ref{mom1}) to close to N-S in channel peak maps (Fig.~\ref{spots_map}). 
We ascribe these differences to the blending of two (spatially) unresolved velocity components.  
In particular, the intensity-integrated velocity in 1$^{st}$ moment maps over pixels of weaker signal, e.g. around the northern component of the radio continuum, could be heavily contaminated by nearby (within the synthesized beam) stronger absorption at different velocities, e.g. coming from the southern component of the radio continuum. 
For these reasons, we consider  the velocity field traced by the velocity-centroid maps more reliable  than the intensity-integrated velocity maps. 

We conclude that the real  velocity gradient in the IRS1 core is close to N-S and has an extent of $\sim$15~\kms. 
 We also note that, since different inversion lines show similar spatial  and 
velocity structure, the high optical depths at lower excitation do not spoil the study of gas kinematics.

\section{Physical Conditions of the Molecular Gas}
\label{phys}

 Using the  parameters measured for the seven \amm~inversion transitions observed in NGC7538~IRS1, 
we can estimate the physical conditions of the \amm\,gas, such as optical depth, kinetic temperature (\tkin), and column density.  
In the assumption of LTE (i.e., the metastable inversion lines are thermalized), 
we can derive the optical depth from fits to hyperfine satellites, 
 the column density from the knowledge of the excitation temperature (\tex) and the opacity,
and the kinetic or rotational temperature (\trot) from column densities of individual metastable levels, 
either with rotational temperature diagrams (RTDs) or ratios of individual transitions \citep{HoTownes83}. 
The methods as well as the formalism adopted to analyze the \amm\,data are described in detail in Appendix~\ref{app}. 

\subsection{Optical Depths}
\label{tau}
 We detected prominent hyperfine pairs of satellite lines spaced $\sim27-31$ \kms\,from the main component, respectively, 
for the (6,6), (7,7), and (9,9) transitions in the core 
and the (6,6) and (7,7) lines in the southern component (Table~\ref{hf_lines}). 
The mere presence of measurable hyperfine satellites in the profiles of highly excited transitions indicates very large optical depths.
Their values can be estimated  using  the theoretical values of the ratio of the satellite line  to the main line strengths, $a_{ms}$, 
reported in Table~\ref{hf_lines} (assuming LTE), 
and comparing them with the measured values (numerically solving  Equation~\ref{tau_eq}).

Toward the core of IRS1 we derive optical depths in the range 10-26 from the (9,9) down to the (6,6) line, 
whereas towards the southern component we estimate 12 and 10 for the (6,6) and (7,7) lines, respectively (see Table~\ref{hf_lines}). 
\citet{Wilson83} estimated 22 and 34 for the (5,5) and (6,6) lines observed with the Effelsberg 100-m telescope, consistent with our JVLA measurements.  
 With spacings of $\sim27-31$ \kms\,and linewidths of $<$10 \kms, these satellite lines can be well resolved from the main line, 
 but since the lines in each  pair of satellites have a frequency difference of only $\sim$0.3-0.4 MHz or 3-5 \kms, 
 they are blended together, making harder an accurate estimate of linewidths.  
 Our estimate of optical depths is affected by this systematics.
 
 Unfortunately, we do not have direct estimates of opacities in the highest excited transitions 
(12,12 for ortho; 10,10; 13,13 and 14,14 for para), for which the satellite lines were too weak. 
For these lines, we used  either the measurements in the lower excitation lines to extrapolate the opacity values in the higher excitation lines or the measured line to continuum ratio. 
These methods are described in Appendix~\ref{tau_app}. 
Using Equations~\ref{tau1} or \ref{tau2}, we obtained much lower values than those directly estimated from the satellites to main lines flux ratios, 
in the range 0.08--0.7 going from the (14,14) to the (10,10) line. 
 An indication that the more highly excited lines are more optically thin comes from the measured values of line to continuum ratios. 
According to Table~\ref{nh3_lines}, this ratio indeed decreases from 0.7 for the (6,6) transition to  0.1 for the (13,13) line. 
Nevertheless, we warn that a lower estimate of optical depths  in these lines
may systematically boost the derived temperatures (see next section).

\subsection{Rotational Temperatures and Column Densities}
\label{td}
 We estimated  rotational temperatures and column densities of the molecular gas using a standard analysis on \amm\,data  \citep[e.g.,][]{HoTownes83}. 
The relevant equations are reported in Appendix~\ref{td_app}.

Ortho- and para-\amm\, have different spin-alignment states, so transitions between para- and ortho-\amm\, are forbidden, and the two behave as separate species \citep{Cheung69}. 
In principle, there could be deviations from their LTE abundance ratio of 1 \citep[e.g.,][]{Goddi11a}. 
Therefore, where possible we estimated  temperatures using  at least two lines of either para or ortho  \amm. 

We consider the central core first. 
We have direct measurements of the opacity only for the lowest-JK lines: (6,6), (7,7), (9,9), 
therefore the most reliable temperature/density estimates come from these lines. 
Using  the (6,6) and the (9,9) lines, 
 we measured a rotational temperature for the absorbing gas, \trot$\sim$290~K, 
and an \amm\,column density over excitation temperature \amm/$T_{ex}\sim5 \times 10^{16}$~cm$^{-2}$/K,   
using both the ratio of the opacities  (Eq.~\ref{tau_ratios}) and the RTD (Eq.~\ref{rtd_eq}). 
When including in the RTD the para (7,7) transitions (assuming an ortho-to-para ratio of 1), 
we obtain a slightly lower \trot$\sim$260~K and a higher  \amm/$T_{ex}\sim7.0 \times 10^{16}$~\cmq/K. 
Allowing for a slightly higher value of ortho-to-para ratio of 1.5 \citep{Goddi11a}, 
we obtain \trot$\sim$280~K and  \amm/\tex$\sim9 \times 10^{16}$~\cmq/K. 
 
For more highly excited lines, we have only poorly constrained lower limits to the opacities. 
Using the three higher para lines (10,10), (13,13), (14,14), we estimate 
 \trot$\sim510$~K and \amm/\tex$\sim3 \times 10^{15}$~\cmq/K. 
Using line ratios,  we obtain \trot$\sim330$~K,   \amm/\tex$ \sim 6-9 \times 10^{15}$~\cmq/K, 
and \trot$\sim460$~K,  \amm/\tex$\sim3-5 \times 10^{15}$~\cmq/K, for the (10,10)-(13,13) and (13,13)-(14,14) pairs, respectively.  
If these highly excited lines were still optically thick, both line ratios and RTDs would overestimate the \trot\,and underestimate the column densities. 
On the other hand, the higher temperatures are not surprising, since  these lines have generally higher level energies (1500-2000 K above the ground), so may be more sensitive to hotter gas, providing rotational temperature closer to the real kinetic temperature of the gas. 
Therefore, if the  (13,13) and (14,14) lines were optically thin, then their ratio would reflect the true temperature,  
indicating the presence of gas with temperatures up to 500~K. 

Adopting \trot=280 K and assuming \tex=\trot, 
we obtain a column density of $\sim1.4-2.5 \times 10^{19}$~\cmq,  averaged on scales $\lesssim$0\pas2. 
This value is an order of magnitude higher than estimated by \citet{Wilson83}, 
who observed the non-metastable (J,K) = (2,1) line with the 100-m telescope 
(therefore their value is averaged on much larger scales, $\sim$40\arcsec).

In the southern component, we detected only the (6,6), (7,7), (9,9), and (10,10) lines, 
and we have direct estimates of opacities only for the (6,6) and (7,7) lines. 
Using an RTD over the first three transitions, we obtain \trot$\sim100$~K and  \amm/$T_{ex}\sim 8.0 \times 10^{16}$~\cmq/K. 
Including also the (10,10) line, we obtain \trot$\sim120$~K and  \amm/$T_{ex}\sim 5.0 \times 10^{16}$~\cmq/K. 
A ratio of the  (6,6) and (9,9) ortho-lines, provide similar values: 
 \trot$\sim120$~K and  \amm/$T_{ex}\sim 3 \times 10^{16}$~\cmq/K. 
 Adopting \trot=110 K and assuming \tex=\trot,  
we estimate a column density of $\sim3-9 \times 10^{18}$~\cmq,  averaged on scales $\lesssim$0\pas2.  
This is consistent with the southern component being at lower temperature and density, 
with respect to the central core, as predicted by \citet{MoscaGoddi14} based on methanol masers.

 As a final note, there are a couple of systematics that may affect the physical condition analysis. 
First, we assume that the different metastable lines fill the same volume. 
Although we are probing relatively small scales, of order of a thousand  AU, 
it is reasonable to expect that the lower excitation lines, such as the  (6,6) doublet, 
fill a larger volume than the higher excitation lines. 
In fact, this seems to be the case, 
with the lowest-JK transitions showing a more uniform N-S distribution in the core  
and the highest-JK transitions concentrating preferably around the two radio continuum peaks,   
as resolved in the VLA A-configuration images  (see Fig.~\ref{spots_map} and discussion in \S~\ref{disk}).  
Second,  the absorption lines are seen against  the continuum, which is optically thick \citep{Sandell09} and therefore its structure may change slightly between 25 and 35 GHz:  
this may potentially affect the line intensities. 
We argue however that both systematics should not affect our estimates of physical conditions. 
In fact our calculations of rotational temperatures   
always consider doublets in adjacent/close energy levels, 
like the (6,6) and (9,9) or the (10,10) and (13,13), but not for example the (7,7) with the (13,13).   
Therefore we can reasonably assume that the transitions used to estimated \trot\,occupy similar volumes. 
Likewise, our analysis involves transitions close in frequency (we never compare transitions at $\sim$25 GHz with transitions at $\sim$ 35 GHz), therefore we do not expect structure changes in the continuum to affect our temperature estimates.

\subsection{Total Mass}
\label{mass}
We can estimate a volume density of molecular gas and a total mass of the core in NGC7538~IRS1, 
by using our estimates of \amm\,column density and assuming a fraction of ammonia with respect to molecular hydrogen. 
For the IRS\,1 core,
we estimate a total column density of molecular gas of $\sim1.4-2.5 \times 10^{26}$~\cmq, 
using $T_{\rm ex}$=280 K and assuming [\amm]/[H$_2$]=10$^{-7}$ 
(typical value for hot-cores - e.g., \citealt{Mauersberger86}). 
In the assumption of spherical distribution of the molecular gas within a radius of 
270~AU\footnote{If we consider the compact core unresolved, the radius is half the beamsize, $\sim$0\pas1 or 270 AU. 
This is approximately the deconvolved value of the radius obtained from Gaussian fitting the core continuum emission.}, 
we estimate a molecular density of $\sim 3.5-6.2 \times 10^{10}$~\cmc\,and a total gas clump mass in the range 19--34 M$_\odot$\footnote{ We explicitly note that the value of [\amm]/[H$_2$] is known within an order of magnitude, therefore the estimates of mass  and molecular density are just order-of-magnitude estimates. 
Likewise, our calculations assume a fixed [\amm]/[H$_2$] ratio in the core, therefore our analysis neglects the effect of chemistry on the location and abundance of ammonia molecules.}. 

For the southern component, using $T_{\rm ex}$=110 K and assuming [\amm]/[H$_2$]=10$^{-7}$, 
we estimate a molecular density of $\sim 0.7-2 \times 10^{10}$~\cmc\,and a gas mass in the range 4--12 M$_\odot$. 
This mass estimate is consistent with the hypothesis of a high-mass YSO forming in the southern component, 
as proposed by \citet{MoscaGoddi14}.

Our analysis indicates an extraordinary concentration of mass in a small area, making IRS1 one of the densest hot-cores known.

\subsection{Comparison with previous studies}

 Assuming thermal equilibrium between the dense gas and dust and 
adopting a temperature of 245 K derived from modeling the CH$_3$CN emission, 
\citet{Qiu11} estimated a gas mass of the core of $20 \pm 12$~\ms\,over about 2\pas5 (or 6000 AU), 
and an average $H_2$ number density of about $10^7$~\cmc\,(from the 1.3~mm dust emission).  
\citet{Beu12} estimates H$_2$ column density of order of $10^{26}$~\cmq, corresponding to an $H_2$ number density of about $2\times 10^9$~\cmc\,averaged over 2000 AU, from which they estimate a total mass of order of 100 \ms. 
Using dust emission ($T_b=219~K$) imaged with the PdBI at comparable resolution to our \amm\,maps, \citet{Beu13} estimates an $H_2$ number density in excess of $10^9$~\cmc\,and a mass of 11--25\ms\,(depending on the assumed dust properties). 
Since we zoom in  further with our \amm\,maps with respect to previous interferometric studies, 
we estimate even higher values for the molecular density (and temperature), 
confirming the special nature of the dense hot core in NGC7538~IRS1.


\section{Discussion}
\label{discussion}

NGC7538~IRS1 is reported to be one of the rare cases of an O-type young star displaying the simultaneous presence of a rotating disk, infalling envelope, and outflow. 
This claim is partly based on interferometric studies generally affected by insufficient linear resolution (of an order of a few thousands AU).   
To complicate matters, the highest angular resolution studies with VLBI of methanol masers and mid-IR imaging disagreed on the disk orientation, 
postulating a rotation axis either NW-SE \citep[e.g.][]{Pes04} or NE-SW \citep[e.g.][]{Buizer05,Surcis11}. 
This led to competing models being proposed to interpret different observations. 

Recently, based on high-angular resolution observations of \met\,masers with VLBI, 
\citet{MoscaGoddi14} presented convincing evidence that IRS1 is actually a multiple system of massive YSOs, 
which solved the controversy aroused in previous models of the region. 
While our \amm\,data do not resolve all the individual YSOs, the multi-transition analysis presented here have enabled us to probe, for the first time, 
the kinematics and physical conditions of the molecular envelope (hosting the multiple) in a wide range of excitation energies, on scales of 500-2000 AU, 
thus complementing the picture provided by VLBI measurements of \met\,masers on smaller scales.  

In the following, we will discuss  
implications of our \amm\,measurements on the multiple system model (\S~\ref{multiple}), 
the evidence of rotation in the (circumbinary) envelope  (\S~\ref{disk}), 
and signatures of infall/outflow in the hot core on different scales (\S~\ref{infall}).

\subsection{A multiple protostellar system in NGC7538~IRS1}
\label{multiple}

With $\sim$0\pas2 resolution, our \amm\,maps clearly resolve IRS1 in 
 the hot core, where the highest excitation \amm~absorption lines originate,  
and a southern component separated by 0\pas4 or 1000 AU from the core, which shows the weakest integrated absorption 
(see Figs.~\ref{mom0}, \ref{mom1}). 
With 0\pas08 resolution, the core component of the radio continuum is further resolved in a northern and southern component, 
0\pas2 or 500 AU apart (see Fig.~\ref{spots_map}). 
The three components of the radio continuum are associated with three different clusters of \met\,masers,  
and \citet{MoscaGoddi14} proposed that they identify a multiple system of massive YSOs: 
IRS1a, IRS1b, and IRS1c.  
IRS1c is associated with the most southern component of the radio continuum and is the least massive of the three. 
Consistently, we do not detect IRS1c in the highest-$JK$ transitions and  
we estimate lower temperatures and densities with respect to the core   (see \S~\ref{phys}). 
The two YSOs IRS1a and IRS1b are forming in the core, have a separation smaller than 500 AU,
and move at velocities of --59.5~\kms\,and --56.4~\kms, as estimated from \met\,maser measurements. 
While these  two objects are not resolved in the intensity and velocity maps of \amm\,(with a resolution of 500 AU), 
 these velocities  are in good agreement with the observed peaks in the spectral line profiles (see Fig.~\ref{spectra-1pan}) 
 and in the  pv-diagrams for different \amm\,transitions (see Figs.~\ref{pv_40} and \ref{pv_991013}).

\begin{figure}
\includegraphics[width=0.5\textwidth]{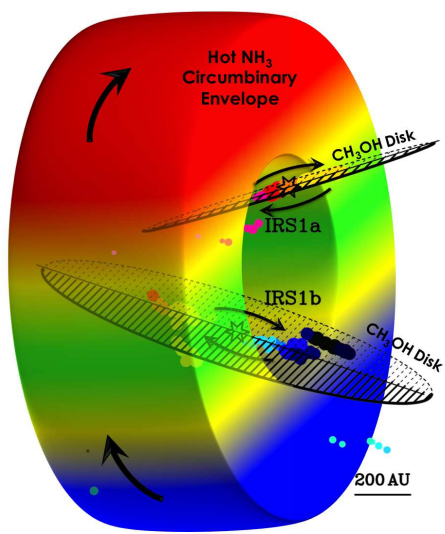}
\caption{Schematic cartoon illustrating several key features of NGC7538~IRS1 that are discussed in the Text. 
6.7 GHz \met\,masers observed with the EVN (filled circles)  identify two high-mass YSOs, IRS1a and IRS1b, surrounded by (quasi-)edge-on disks. 
The black ellipses represent the disk planes, with the solid lines indicating the near-side and the dashed lines the far-side of the disks. 
These star-disk systems reside in the inner cavity of the circumbinary envelope (probed by hot ammonia), cleared up by the orbital motion of the binary members. 
The cartoon shows only the central section of such a molecular envelope (supposedly spherical).  
The gas in the envelope may feed the circumstellar disks of IRS1a/b (this is however of too high-order complexity to be shown in the cartoon).
Colors denote line-of-sight velocities, with blue indicating blue-shifted emission and red indicating red-shifted emission, 
while green represents systemic velocity.  
Material in both the disks and the envelope is rotating, in the sense indicated by the black arrows.}
\label{cartoon}
\end{figure}

If two YSOs IRS1a and IRS1b are forming in the core, we expect gas motion both {\it between} the two YSOs (in the natal core) 
and {\it around} each individual YSO (in accretion disks). 
 \citet{MoscaGoddi14} presented compelling evidence that both IRS1a and IRS1b are surrounded by rotating disks. 
Since their size of 500 AU, as inferred from the masers, is comparable to the linear resolution of our \amm\,observations, 
we do not resolve individual disks. 
\amm\,probes instead dense and hot molecular material rotating in the envelope that is likely feeding the two accretion disks. 
 A cartoon schematizing this model is presented in Figure~\ref{cartoon}. 
We investigate the kinematic properties of the \amm\,envelope in the next section.

\subsection{A rotating (circumbinary) envelope in NGC7538~IRS1}
\label{disk}

The compact structure of the \amm\,core shows a clear velocity gradient, in both the velocity field maps (Fig.~\ref{mom1}) 
and the position-velocity plots (Figs~\ref{pv_40} and \ref{pv_991013}), 
which may be indicative of rotation in the envelope surrounding the two high-mass YSOs IRS1a and IRS1b.  
Since the linear size of this envelope is $\sim$500 AU, 
comparable with the separation between IRS1a and IRS1b and with their disk sizes \citep{MoscaGoddi14}, 
the hot gas surrounding the two YSOs is blended in our \amm\,maps.
While the observed velocity gradient  in the $1^{}st$ moment maps may be affected by the blending of the two objects,    
the velocity field from plots of (Gaussian-fitted) channel peak positions is not, 
since the two YSOs move at different velocities. 
 Therefore, we consider the velocity field traced by the velocity-centroid maps more reliable than the intensity-integrated velocity maps 
(as already argued in \S~\ref{spots}), and, accordingly, we will base the following discussion on the former.

A few considerations are in order. 

First, the direction of the  velocity gradient is close to N--S,  
i.e. the direction along the line connecting the two continuum peaks or YSOs IRS1a and IRS1b. 
We note that the \amm\,velocity-centroids do not trace material in the individual rotating disks (which are unresolved in our maps).  
Rather, each centroid represents the average velocity of the molecular material that is orbiting around the two YSOs inside the natal molecular core. 
Thefore, the  velocity gradient in the \amm\,gas between the two continuum peaks implies that 
the molecular gas inside the core rotates on the orbital plane of the binary. 

Second, there is no  steepening of the velocity gradient with the increase of the excitation energy.  
This is inconsistent with molecular material orbiting around one single central YSO, 
since one would naturally expect higher-excitation energy lines to probe inner radii. 
But this is still consistent with two ``heating centers'', e.g. one at the northern and one at the southern edge of the core. 
In fact, in this case one would expect different transitions to trace a similar N-S distribution with a size given by the separation of the two YSOs, 
and to observe the hottest material around each YSO (i.e. around the northern and southern continuum peaks), 
and the colder material in between.  
This is indeed what we observe, with the highest-JK transitions showing weaker absorption in between the two radio continuum peaks, 
and the lowest-JK transitions showing a more uniform distribution  (see Fig.~\ref{spots_map}). 
This explains also why the pv-diagrams from different transitions probe similar regions (Figure~\ref{pv_991013}).

A final interesting aspect worth discussing is the velocity profile. 
While the observed velocity gradient   is consistent with rotation, 
     the structural asymmetry in different quadrants shown in the overlay of pv-diagrams from different transitions   (Figure~\ref{pv_991013}), 
     indicates a complex (non-Keplerian) profile. 
This is not surprising, since the \amm\,absorption in IRS1 is not tracing a rotating disk but rather a flattened rotating envelope, for which we do not expect a Keplerian profile.
      Nevertheless, as an order of magnitude, 
we could calculate the rotationally supported binding mass $M_{\rm{bind}}$ for the core in NGC7538~IRS1,  
assuming equilibrium between the gravitational and rotational force at the outer envelope radius:
$ M_{\rm bind}  =  V^2_{\rm rot} \ R \ / \ G \ sin^2(i)  $, \
where  $R$ is the envelope radius, V$_{\rm rot}$ is rotational velocity, 
and $i$ is the inclination angle between the rotation axis and the line of sight
($i=90^{\circ}$ for an edge-on system).
Taking as radius half the separation between the continuum peaks, $\sim$0\pas1 or 270 AU, 
and as rotation velocity the total range in the velocity-centroid maps divided by 2, $V_{\rm rot}\sim$7.5\,km\,s$^{-1}$, 
 the enclosed mass would be $\sim 42$/sin$^2(i)$\,M$_{\odot}$. 
 This mass value agrees well with the total (stars + gas)  mass estimated from the VLBI measurements of CH$_3$OH masers (25 + 16 = 41~\ms) , for a system close to edge-on.
This total mass is also higher than the  gas mass from \amm\,column density, 19--34~\ms\,(\S~\ref{mass}), indicating that substantial mass is contained in the central stars.

We conclude that, although we cannot confirm the presence of protostellar disks around individual YSOs,  
the \amm\,images presented here still indicate the presence of a rotating core of size $\sim$500~AU, 
feeding the two smaller disks observed in \met\,masers, as schematized in Figure~\ref{cartoon}.

\subsection{Is the \amm\,core infalling or outflowing?}
\label{infall}

 Previous interferometric studies on NGC7538~IRS1, 
 have provided a contradictory picture regarding the dominant motion in the core, 
 depending on the  spatial scales probed.

Lower-resolution  NH$_3$ (1,1), (2,2) observations (7--10\arcsec~resolution) revealed P-Cygni profiles
with \emph{blueshifted} absorption (--60~\kms) toward IRS 1 \citep[e.g.,][]{Keto91,Zheng01}. 
Since the NH$_3$ absorption is optically thick and therefore 
presumably traces gas residing in outer layers in the foreground of the HII region, 
the absorption was attributed to outflowing gas toward the observer  \citep[e.g.,][]{Zheng01}. 
 Subsequent higher resolution (0\farcs7-3\arcsec) interferometric studies  \citep{Qiu11,Beu12,Zhu13}, 
revealed  {\it inverse} P-Cygni profiles with {\it redshifted} absorption (from --54.6 to --57~\kms) 
in (sub)mm molecular lines from different species, like 
 C$^{18}$O, SO, HCN, HNCO, HCO$^+$, OCS, CH$_3$OH, and CH$_3$CN 
 (see for ex., Figure~8 from  \citealt{Qiu11}). 
  In these lines, the absorption comes from dense gas lying in the foreground of the 
bright continuum emission from the HII region and moving toward the source (away from the observer); 
the emission comes instead from dense gas residing in the background of the HII region and moving toward the observer. 

How can we reconcile the (contrasting) signatures on small and large scales? 
As discussed in Section~\ref{multiple}, the core of IRS1 hosts two compact objects, 
IRS1a, the dominant object at systemic velocity (--59.3~\kms), and IRS1b, a lower mass object with redshifted velocity (--56.4~\kms); 
a third object, IRS1c, is associated with the southern component of the radio continuum and \amm\,absorption, 
and is moving with blueshifted velocity  (--59.8~\kms). 
In Section~\ref{res}, based on spectral profiles and pv-diagrams, we demonstrated that the lower-excitation lines tracked here, 
from (6,6) to (9,9), are dominated by the object at --59.3~\kms, with some weak emission contributed by the southern blueshifted YSO IRS1c. 
 With higher excitation energy, the redshifted component at 56.4~\kms\,becomes more comparable to the main component 
 (e.g., for the 13,13 line both components show similar strengths; see Figure~\ref{pv_40}), whereas the blueshifted southernmost  component disappears. 
Based on this finding, we conclude that the blueshifted absorption observed on large scales in the  NH$_3$ (1,1), (2,2) inversion lines 
is not tracing outflowing gas, but  most probably material associated with IRS1a and IRS1c. 
On the other hand, (sub)mm lines of higher excitation would be sensitive to all (blueshifted and redshifted) velocity components. 
The presence of both blueshifted emission and redshifted absorption in these (sub)mm lines \citep{Qiu11,Beu13}, indicates that the infall signature is real.

If the redshifted absorption that we observe in the highly-excited \amm\,inversion lines  (with $E_l$ up to 1950 K) were due to infall, 
this would indicate ongoing accretion of very hot and dense gas across the HII region within a radius of only 270 AU. 
While we clearly detect the absorption, our observations are not however sensitive enough to detect highly excited \amm\,in emission. 
Therefore, without a full inverse P-Cygni profile we do not have a reference velocity, 
and we cannot conclude if the highly excited \amm\,gas is infalling or outflowing. 
Nevertheless, recent observations of the HCN (4-3) $\nu_2=1$ line ($E_{\mathrm {up}}\sim1000$ K) with the PdBI, 
with an angular resolution  of 0\pas2 (comparable with our JVLA measurements), 
revealed a clear inverse P-Cygni profile tracing infalling gas at a rate of $\approx\ 3.6 \times10^{-3} \ \rm M_\odot$ yr$^{-1}$  \citep{Beu13}. 
Therefore,  it may be the case that the \amm\,gas in the core is infalling onto the multiple protostellar system at a similar rate.

\section{Summary}
We imaged the core of NGC7538~IRS1 with the JVLA at a 0\pas2 resolution in seven metastable inversion transitions of ammonia, which are emitted from doublet levels from about 400 K up to 1950 K above the ground state. 
We conclude the following:

\begin{enumerate}[1.]

\vspace{0.2 cm} 
\item
The highly-excited \amm\,inversion transitions are observed in absorption against the strong HC-HII region associated with  NGC7538~IRS1. 
With the 500 AU linear resolution, we resolve the elongated North-South \amm\,structure into two compact components, the main core and a southern component.

\vspace{0.2 cm} 
\item 
Based on intensity-weighted velocity  (or first moment) maps,  the \amm\ inversion lines show an apparent  velocity gradient oriented NE--SW, 
with a total extent of 9~\kms.   
This velocity gradient has been already observed in other molecular transitions and has been interpreted 
as a rotating disk perpendicular to a known CO outflow. 
Examining the pv-diagrams, however, we do not find a Keplerian profile, 
nor a steepening of the gradient with  increasing line excitation energy
(one would expect warmer gas closer to the central YSO to move faster).  
Although this argues  against the disk interpretation, 
we demonstrate that the velocity gradient still traces gas rotation in the molecular core (rather than in a disk; see item 4).

\vspace{0.2 cm} 
\item
VLBI measurements of \met\,masers recently provided compelling evidence that IRS1 is forming a triple system of high-mass YSOs:  IRS1a, IRS1b, and IRS1c. 
IRS1c, with a line-of-sight velocity around --60~\kms, is associated with the southernmost  component of the radio continuum, 
separated by a 1000 AU or 0\pas4 from the core, 
and is not detected in the highest-$JK$ transitions,  indicating lower-excitation temperatures. 
The two YSOs IRS1a and IRS1b, embedded in the hot molecular core, 
are associated with two resolved components of the radio continuum imaged with 0\pas08 (or 200 AU) resolution, 
have a separation smaller than 500 AU or 0\pas2,
and move at velocities around --59~\kms\, and --56~\kms, respectively.
While these  two objects are not resolved in the intensity and velocity maps of \amm, 
with a resolution of 500 AU, 
 these velocities  are in good agreement with the observed peaks in the spectral line profiles 
 and in the  pv-diagrams for different \amm\,transitions. 
JVLA observations in the A-configuration are needed to resolve the binary system in the core with a separation of 500 AU.

\vspace{0.2 cm} 
\item
 Resolved images of \amm\ absorption peak positions in individual spectral channels  (fitted with Gaussians) 
identify a  velocity gradient close to N--S, 
i.e. the direction along the line connecting the two components of the radio continuum separated by 500 AU or 0\pas2. 
This reflects the global rotation of the natal massive core, in the orbital plane of the binary composed by the two high-mass YSOs IRS1a and IRS1b. 
With an apparent extent of $\sim$~15~\kms, this  velocity gradient corresponds to an enclosed  mass in the core of $\sim42$~\ms\,
(assuming the \amm\,gas is rotating in centrifugal equilibrium), 
in good agreement with the total dynamical mass estimated from  VLBI measurements of CH$_3$OH masers  (41~\ms). 

\vspace{0.2 cm} 
\item 
From  simple LTE analysis, the molecular gas in the core has a temperature of 280~K, with a potential hotter component up to 500 K. 
The H$_2$ density is over $10^{10}$~\cmc\,(assuming [\amm]/[H$_2$]=10$^{-7}$), making NGC7538~IRS1 the densest hot core known.  
We also estimate a gas mass from \amm\,of 19--34~\ms\,for the core, 
 which is lower than  the "total"  (stars + gas) dynamical masses estimated from \amm\,and \met\,masers.  
 The southern component has a lower  temperature, around hundred K, but similar (though lower) molecular gas density.  

\end{enumerate}

This study on NGC7538~IRS1, demonstrates that high-angular (i.e. a few tenths of arcseconds) resolution imaging of high-excitation lines of \amm\,
at $\sim$1~cm wavelengths are well suited to study the kinematics and physical conditions of the hottest and densest molecular gas in the vicinity of the central high-mass YSO(s), 
that is in accretion disks and the innermost regions of circumstellar envelopes. 
Our observational program at the JVLA aims to apply the analysis presented in this paper to a sizable sample 
of very luminous Galactic hot molecular cores hosting O-type YSOs. 
The ultimate goal is to characterize observationally the mass-accretion process  in the most massive young stars. 
\\


\appendix

\section{Methods to estimate the physical conditions of the \amm\,gas}
\label{app}
In this Appendix, we describe the methods and formalism we adopted to estimate the physical conditions, such as optical depth, rotational temperature, and column density,  of the \amm\, gas.

\subsection{Optical Depth Analysis}
\label{tau_app}

 When both the main line and the hyperfine satellites are detected, 
the line opacity of one specific inversion transition can be estimated numerically by comparing the measured and theoretical ratios 
of the satellite line  to the main line strengths \citep[e.g.,][]{HoTownes83}: 
\begin{equation}
\label{tau_eq}
\frac{F_{main}}{F_{sat}} = \frac{1-e^{-\tau}}{1-e^{-a_{ms}\tau}} 
\end{equation} 
  (the theoretical values $a_{ms}$ are reported in Table~\ref{hf_lines}).
  
This method assumes that the hyperfine satellites are  optically thin,  
a  reasonable assumption since the relative line strengths of the satellite lines 
are a very small fraction of the main line intensity ($<$1\% for the (6,6) and upper transitions).  

For the highest excited transitions, the satellite lines are too weak to be directly detected.  
For those lines,  two indirect methods  can be used for estimating the optical depths. 
In the cases where $\tau(J_1,K_1)$ can be measured for an adjacent lower state, 
 $\tau(J_2,K_2)$ for the  higher state can be deduced from the ratio of the observed brightness temperatures or integrated fluxes  \citep[e.g.,][]{HoTownes83}: 
\begin{equation}
\label{tau1}
\tau(J_2,K_2) = \log \left[ 1 - \frac{ (F(J_2,K_2) \nu^2_1 }{ (F(J_1,K_1) \nu^2_2 } \times (1 - e^{-\tau(J_1,K_1)} ) \right]
\end{equation} 

In alternative, the optical depth can be computed from the measured line to continuum ratio using the relation: 
\begin{equation}
\label{tau2}
\tau(J,K) = - \log \left[ 1 - \frac{F(J,K)}{F_{cont}} \right]
\end{equation} 

where $F(J,K)$ is line peak flux and $F_{cont}$ is the continuum flux integrated in the area of the absorption line. 
 Here, we assume that the source "covering factor" (the fraction of the continuum covered by the \amm\,absorbing cloud along the line-of-sight) is close to unity since the radio continuum emission is compact towards the core.  
This method works best if the optical depths are low. 
In case of high opacities, however, $F(J,K)/F_{cont}$ is close to unity, therefore the exact value of $\tau$ becomes very uncertain. 

Both methods may provide lower estimates of the optical depths, which in turn
may systematically boost the derived temperatures.

\subsection{Temperature and column density analysis}
\label{td_app} 

The \amm\,populations can be described in terms of different temperatures.  
The excitation temperature, \tex, characterizes 
the relative populations of the lower and upper levels of a (J,K) inversion doublet, 
according to the Boltzmann distribution: ${N_u}/{N_l} =g_u/g_l  e^{(E_u-E_l) / T_{ex}}$.   
The rotational temperature, \trot, characterizes the ratio of the populations of two (J,K) doublets. 
Since the metastable populations can only be exchanged by collisions (radiative transitions are forbidden between K levels), 
characterized by the kinetic temperature, \tkin, there is a direct relation between \trot\,and \tkin, 
with \trot\,close but generally lower than \tkin. 
When collisions dominate the excitation and radiative processes are unimportant, 
the population reaches local thermal equilibrium (LTE), and it is customary to assume  \tex=\trot=\tkin. 
This assumption is generally valid for metastable inversion lines \citep[e.g.,][]{Walmsley83},  
and in general for hot molecular cores, which have typical densities  $>10^7$~\cmc.   
In the following analysis, we therefore assume that a single excitation temperature (equal to a kinetic temperature), 
describes all the observed metastable transitions. 

The column density of one optically thick \amm\,inversion line in absorption may be found from the line optical depth: 

\begin{equation}
\label{N_tau}
 N(J,K)  = 1.6 \times 10^{14} \ \frac{J(J+1)}{K^2} \ \frac{\Delta v}{\nu} \ \tau \ T_{ex}
\end{equation} 
where 
$N(J,K)$ is in cm$^{-2}$,
$\Delta v$ is the linewidth (in \kms), 
$\nu$ is the transition frequency (in GHz), 
\tex\,is the excitation temperature (in K), 
and $\tau$ is the line opacity \citep[e.g.,][]{Huettemeister95}. 

In order to find the total column density, we assume that all the energy levels  are populated according to a Boltzmann distribution, 
characterized by a single \trot. 
The total column density of \amm\,can then be expressed as: 
\begin{equation}
\label{eqboltz1}
N = \frac{N(J,K)}{g(J,K)} \ Q(T_{rot}) \ e^{E_{l}/T_{rot}}  
\end{equation} 
where 
$g(J,K)=g_{op}\ (2J+1)$ are the statistical weights ($g_{op}=1$ for para- and  $g_{op}=2$ for ortho-transitions), 
$Q(T_{rot})$ is the partition function, and $E_l$ is the lower state energy of the (J,K) transition (in K). 
Therefore, assuming Boltzmann statistics, we need to estimate \trot\,in order to estimate $N$. 
We use two different methods. 

The rotational temperature between two different levels can be derived in terms of the ratio of their optical depths \citep[e.g.,][]{HoTownes83}:
\begin{eqnarray}
\label{tau_ratios}
\frac{\tau(J_2 , K_2)}{\tau(J_1,K_1)}  =   \nonumber  \\ 
= \frac{ \nu^2(J_2,K_2) }{ \nu^2(J_1,K_1)} \frac{\Delta v(J_1,K_1)}{\Delta v(J_2,K_2)} \frac{T_{ex}(J_1,K_1)}{T_{ex}(J_2,K_2)}  \frac{ \mu^2(J_2,K_2) }{ \mu^2(J_1,K_1)}  \frac{ g(J_2,K_2) }{g(J_1,K_1)} \\
\times  \ e^{ (E_2 - E_1) / T_{rot} } \nonumber
\end{eqnarray}
where $ \mu^2(J,K)= \mu^2 \frac{K}{[J(J+1)]}$ are the dipole matrix elements, 
$E_1$ and $E_2$ are the lower state energies of the  ($J_1,K_1$) and ($J_2,K_2$) doublets, 
 and $T_{rot}$ is the rotational temperature between them. 
 Once \trot\,is known, we can use Equations~\ref{N_tau} and \ref{eqboltz1} to derive an \amm\,column density.

When more than two transitions are observed, we can use RTDs, where the rotational temperature and the column density  are measured simultaneously. 
Rearranging Equation~\ref{eqboltz1} and using Equation~\ref{N_tau} for the column density $N(J,K)$, we have:  

\begin{eqnarray}
\label{N}
 1.6 \times 10^{14} \frac{J(J+1)}{K^2}\frac{1}{g_{op}*(2J+1)} \frac{\Delta v}{\nu} \tau = \frac{N}{T_{ex} Q(T_{kin})}    \ e^{-E_{l}/T_{kin}}  
\end{eqnarray}

Taking the natural logarithm of both sides, we find: 
\begin{eqnarray}
\label{rtd_eq}
\log{\left[ 1.6 \times 10^{14} \frac{J(J+1)}{K^2}\frac{1}{g_{op}*(2J+1)} \frac{\Delta v}{\nu} \tau \right] }  \nonumber \\ 
= \log{\frac{N}{T_{ex} Q(T_{rot})}} - \frac{E_l}{T_{rot}}  
\end{eqnarray}

Equation~\ref{rtd_eq} shows that the logarithm of the left member is a linear function of $E_l$ 
(if all transitions have the same \tex), 
with slope $-1/T_{rot}$ and intercept $log\frac{N}{T_{ex} Q(T_{rot})} $ at  $E_l=0$. 
We can then determine \trot\, and $N$ from a least squares fit of lower state energy  to the optical depth for different transitions in log space.  \\


\begin{acknowledgements}
We thank the anonymous referee for a constructive report. 
We are grateful to Xing Lu and A. Sanchez-Monge for useful discussion and help on different spectral line fitting methods.  
These data were obtained under JVLA program 12A-274.
\end{acknowledgements}


\bibliographystyle{aa}
\bibliography{biblio}  

\end{document}